\documentclass[journal]{IEEEtran}
\usepackage{cite}
\usepackage{amsmath,amssymb,amsfonts}
\usepackage{algorithm,algorithmic}
\usepackage{graphicx,subfig,color, upgreek}
\usepackage{array,multirow}

\begin{document}
\title{An Efficient Algorithm for Spatial-Spectral Partial Volume Compartment Mapping with Applications to Multicomponent Diffusion and Relaxation MRI}
\author{Yunsong Liu, \IEEEmembership{\small Member, IEEE}, Debdut Mandal, Congyu Liao, Kawin Setsompop, Justin P. Haldar, {\IEEEmembership{\small  Senior Member, IEEE}}
\thanks{This work was supported in part by NIH grants R01-MH116173,  R01-NS074980, and R56-EB034349, a USC Viterbi Graduate Fellowship, and resources from the USC Center for Advanced Research Computing.}
\thanks{Y. Liu, D. Mandal, and J. Haldar are with the Signal and Image Processing Institute, Ming Hsieh Department of Electrical and Computer Engineering, University of Southern California, Los Angeles, CA, 90089 USA (e-mail: yunsongl@usc.edu, debdutma@usc.edu and jhaldar@usc.edu). }
\thanks{C. Liao and K. Setsompop are with the Departments of Radiology and Electrical Engineering, Stanford University, Stanford, CA 94305, USA (e-mail: cyliao@stanford.edu, kawins@stanford.edu)}}

\maketitle 
 
\begin{abstract}
We introduce a new algorithm to solve a regularized spatial-spectral image estimation problem.  Our approach is based on the linearized alternating directions method of multipliers (LADMM), which is a variation of the popular ADMM algorithm.   Although LADMM has existed for some time, it has not been very widely used in the computational imaging literature.  This is in part because there are many possible ways of mapping LADMM to a specific optimization problem,  and it is nontrivial to find a computationally efficient implementation out of the many competing alternatives.  We believe that our proposed implementation represents the first application of LADMM to the type of optimization problem considered in this work (involving a linear-mixture forward model, spatial regularization, and nonnegativity constraints).    We evaluate our algorithm in a variety of multiparametric MRI partial volume mapping scenarios (diffusion-relaxation, relaxation-relaxation, relaxometry, and fingerprinting), where we  consistently observe substantial ($\sim$3$\times$-50$\times$) speed improvements.  We expect this to reduce barriers to using spatially-regularized partial volume compartment mapping methods.  Further, the considerable improvements we observed also suggest the potential value of considering LADMM  for a broader set of computational imaging problems.

\end{abstract}


\section{Introduction}
\label{sec:introduction}
Partial volume effects occur in virtually all medical imaging experiments, and result from the fact that  biological tissues contain high-resolution features that are smaller than the achievable spatial resolution. This means that the signal observed from a macroscopic image voxel will represent a mixture of different microscopic sub-voxel tissue compartments. 

This paper is focused on the problem of unmixing  sub-voxel tissue compartments. Methods that can solve such problems are significant because they provide information about  tissue structure that is too small to be directly observed with conventional imaging.

To robustly solve the partial volume unmixing problem, it is important  to acquire  imaging data that is sensitive to the differences between different sub-voxel compartments. One common way to achieve this is to acquire a series of images with different experimental contrast settings, where the contrast variations will affect the sub-voxel compartments in distinct ways, following a known parametric  model. For a finite number of components and assuming the mixture is linear, this leads to the data acquisition model
\begin{equation}
 m_p(\mathbf{r}_n) = \sum_{q=1}^Q b(\boldsymbol{\uptheta}_p, \boldsymbol{\upgamma}_q) f_q(\mathbf{r}_n) + \eta_p(\mathbf{r}_n)\label{eq:model}
\end{equation}
for $p=1,\ldots,P$ and $n=1,\ldots,N$, where $P$ is the total number of measured images in the image sequence; the vector $\mathbf{r}$ denotes spatial location  and it is assumed that we observe $N$ voxels at spatial positions $\mathbf{r}_n$, $n=1,\ldots,N$; $m_p(\mathbf{r})$ is the $p$th measured image; $\eta_p(\mathbf{r})$ is the noise within the $p$th measured image; $\boldsymbol{\uptheta}_p$ are the contrast-related experimental parameters for the $p$th measured image; $Q$ is the number of compartments; $\boldsymbol{\upgamma}_q$ are the contrast-related tissue parameters for the $q$th compartment; $f_q(\mathbf{r})$ is a spatial map  representing the  contribution\footnote{Please note that while some of the literature on partial volume compartment mapping relies on fraction-based representations where the $f_q(\mathbf{r})$ are normalized maps that are required to sum to 1, we do not use such an approach in our work.  We prefer to leave the compartments unnormalized, as fractional normalization can introduce ambiguities that confound interpretation.  For example, it is difficult to identify whether a change in a fraction map is due to a change in the current compartment (the numerator) or a change in other compartments (the denominator), and fraction maps can also hide potentially interesting changes in total signal intensity that do not change the relative volume fractions.  Non-fractional representations do not suffer from these ambiguities, and are also more natural for the type of smoothness-based spatial constraints used later in the paper. } of the $q$th sub-voxel tissue compartment to the total signal; and $b(\boldsymbol{\uptheta},\boldsymbol{\upgamma})$ is a model of the ideal data that would be expected from pure tissue with parameters $\boldsymbol{\upgamma}$ under experimental parameters $\boldsymbol{\uptheta}$.  Given a model of this form, the contrast parameters $\boldsymbol{\upgamma}_q$ and spatial maps $f_q(\mathbf{r})$ for each sub-voxel compartment can be estimated from the measured data by solving an appropriate inverse problem.  

Our  interest in this type of mixture modeling stems primarily from our own recent work on  multiparametric correlation-spectroscopic imaging in MRI \cite{Daeun2017,Daeun2020} (involving multidimensional diffusion, $T_1$-relaxation, and/or $T_2$-relaxation encoding -- see also \cite{McGivney2018,deshmane2019,Tang2018,nagtegaal2020,slator2021,benjamini2020} for related  work from other groups). However, multicomponent mixture models are also widely used in a range of other experimental paradigms including dynamic PET  \cite{osullivan1993}, dynamic contrast-enhanced MRI and CT \cite{ingrisch2013},  MRI relaxometry \cite{Does2018}, and diffusion MRI \cite{Alexander2019}, to name just a few.

In many applications, it is common for  the compartmental signal model $b(\boldsymbol{\uptheta}, \boldsymbol{\upgamma})$ to resemble some form of exponential decay, which emerges as a consequence of the imaging physics.  As a result, the unmixing problem reduces to estimating the parameters of  a multiexponential decay model.  This kind of multiexponential estimation  problem (sometimes informally called an ``inverse Laplace transform'') appears frequently in the physical sciences, and has been studied for hundreds of years by the mathematics and physics communities \cite{istratov1999}.  Unfortunately, it has also been understood for hundreds of years that this is a highly ill-posed problem. This means that there can be  many different parameter choices that all fit the measured data similarly well, and also that estimated solutions can be highly sensitive to small noise perturbations.  

Different strategies have been developed to tackle this fundamental ill-posedness.   From a data acquisition perspective, it has been demonstrated that multidimensional multiparametric MRI acquisitions (e.g., encoding $T_1$-$T_2$ or diffusion-$T_2$ parameters simultaneously and nonseparably) can lead to inverse problems that are substantially better posed than lower-dimensional single-parameter acquisitions (e.g., that encode $T_1$, $T_2$, or diffusion parameters alone) \cite{Daeun2020,celik2013}.  From an estimation perspective, it has proven useful to introduce various types of constraints on the solution to the inverse problem.  While many different types of constraints have been introduced over the decades \cite{Kroeker1986, Whittall1989, venkataramanan2002,  bai2015, benjamini2016, Daeun2017, McGivney2018, Tang2018,  deshmane2019,  nagtegaal2020, benjamini2020, Daeun2020, slator2021, lin2014, Hwang2009, Kumar2012,  Labadie2014, Kumar2018, Zimmermann2019}, this paper will focus exclusively on methods that assume spatial regularity of the compartmental spatial maps $f_q(\mathbf{r})$ \cite{Hwang2009, Kumar2012, lin2014, Labadie2014, Daeun2017, Kumar2018, Zimmermann2019, Daeun2020}.  It has been shown theoretically that the use of spatial constraints can dramatically reduce the ill-posedness of the inverse problem in both one-dimensional \cite{lin2014} and multidimensional \cite{Daeun2020} acquisition scenarios compared to the traditional approach in which the mixture model is solved independently for each voxel.

While multidimensional multiparametric acquisition approaches and spatially-regularized compartment estimation approaches can both greatly improve the quality of partial volume unmixing, these approaches generally lead to increased memory and computational complexity requirements.  Multidimensional multiparametric acquisition naturally leads to increased computational complexity because of the curse of dimensionality.  Meanwhile, spatial regularization methods require increased computational complexity because it becomes necessary to estimate the model for multiple voxels simultaneously instead of solving for each voxel independently.  This increased computational complexity can be quite burdensome, and we are aware of only a few examples where spatial regularization has been used to estimate large-scale spatial maps $f_q(\mathbf{r})$, $q=1,\ldots,Q$ \cite{lin2014,Daeun2017,Daeun2020, Zimmermann2019}, with other spatially-constrained methods opting instead to either only estimate small local image patches or using other heuristics to avoid the computational complexity of the full spatially-regularized inverse problem \cite{Hwang2009, Kumar2012, Labadie2014, Kumar2018}. 

Recent methods for solving large-scale spatially-regularized partial volume compartment mapping problems \cite{Daeun2017, Daeun2020, Zimmermann2019} have all made use of algorithms based on the alternating direction method of multipliers (ADMM) \cite{Boyd2011}.  While there are some differences in formulation and  implementation, all of these algorithms introduce multiple sets of auxiliary variables, which, through clever use of variable-splitting, result in simplified subproblems that are each solved efficiently.  These algorithms produce good results, although they often converge slowly and consume large amounts of memory. 
	
In this paper, we explore a variant of ADMM called linearized ADMM (LADMM) \cite{Zhang2011,He2020} (also known as generalized\cite{Deng2016} and proximal \cite{Fazel2013,Tao2020} ADMM) to alleviate the issues with ADMM in this setting.  In practice, there are many potential ways to map the general LADMM framework to a specific optimization problem, and it can require nontrivial problem-specific insights to identify an approach that will yield improved computational efficiency.  Our proposed approach, obtained after substantial experimentation with different options, is based on two insights: first, that we can save substantial amounts of memory by reducing the number of auxiliary variables; and second, that we can also potentially improve convergence speed by reducing the number  and complexity of  subproblems that need to be solved compared to previous ADMM methods. This results in a smaller number of subproblems that capture more of the structure from the original inverse problem compared to the previous algorithms, enabling faster convergence with much less memory usage.  A preliminary account of portions of this work was previously presented in a recent conference \cite{liu2022}. 

This paper is organized as follows.  In Sec.~\ref{sec:form}, we provide a detailed description of the optimization problem that we are interested in solving, and review  existing algorithmic approaches.  Then in Sec.~\ref{sec:proposed}, we review ADMM and LADMM and describe how we adapt LADMM  to the problem at hand.  In Sec.~\ref{sec:results}, we compare our new algorithm against ADMM in a range of MRI applications, including multicomponent diffusion-$T_2$ relaxation correlation spectroscopic imaging (DR-CSI) \cite{Daeun2017}, multicomponent $T_1$-$T_2$ relaxation correlation spectroscopic imaging (RR-CSI) with both inversion recovery multi-echo spin-echo (IR-MSE) \cite{Daeun2020} and magnetic resonance fingerprinting (MRF)  \cite{kim2019, liu2022} acquisitions, and multicomponent $T_2$ relaxometry \cite{Whittall1989, Kroeker1986, Does2018}.  Discussion and conclusions are presented in Secs.~\ref{sec:disc} and \ref{sec:conc}, respectively.

\section{Problem Formulation and Existing Methods}\label{sec:form}
\subsection{Formal Description of Optimization Problem}
While there are many different multicomponent mixture modeling approaches that can emerge from Eq.~\eqref{eq:model}, our attention in this work is restricted to the class of  ``spectral'' or ``continuum'' models \cite{venkataramanan2002, slator2021, Daeun2020,provencher1982}.  In this setting, rather than assuming that the number of compartments $Q$ is a small finite number corresponding to a discrete set of tissue parameters $\boldsymbol{\upgamma}_q$ as in Eq.~\eqref{eq:model}, it is instead assumed that the measured data results from a continuous mixture of contributions arising from a potentially-infinite set of tissue parameter values.  This leads to the integral equation
\begin{equation}
 m_p(\mathbf{r}_n) = \int  b(\boldsymbol{\uptheta}_p, \boldsymbol{\upgamma}) f(\boldsymbol{\upgamma},\mathbf{r}_n)d\boldsymbol{\upgamma} + \eta_p(\mathbf{r}_n), \label{eq:cont}
\end{equation}
for $p=1,\ldots,P$ and $n=1,\ldots,N$, 
where $f(\boldsymbol{\upgamma},\mathbf{r})$ is a high-dimensional ``spectroscopic image'' that includes both spatial dimensions $\mathbf{r}$ and spectral dimensions $\boldsymbol{\upgamma}$.  For practical implementation, it is common to discretize the spectroscopic image at a finite number of $Q$ spectral positions $\boldsymbol{\upgamma}_q$, $q=1,\ldots,Q$, which results in
\begin{equation}
 m_p(\mathbf{r}_n) = \sum_{q=1}^Q w_q   b(\boldsymbol{\uptheta}_p, \boldsymbol{\upgamma}_q) f(\boldsymbol{\upgamma}_q,\mathbf{r}_n) + \eta_p(\mathbf{r}_n), \label{eq:cont2}
\end{equation}
for $p=1,\ldots,P$ and $n=1,\ldots,N$,
where $w_q$ are the density normalization coefficients (i.e., numerical quadrature weights) required for accurate approximation of the continuous integral using a finite discrete sum.

Assuming that we work with real-valued images,\footnote{ Note that MRI images are generally complex-valued, and this assumption is not strictly necessary.  However, it is common practice to work with real-valued magnitude images in the MRI literature on partial volume mixture modeling, since it simplifies the use of nonnegativity constraints (which are powerful and widely-used in this setting) when all compartments from the same voxel are assumed to share the same phase  \cite{Daeun2017,Daeun2020}. Note that when phase is relevant, it is also possible to obtain real-valued images from complex-valued images through appropriate phase modeling. }  Eq.~\eqref{eq:cont2} can be conveniently expressed in matrix-vector form as
\begin{equation}
 \mathbf{m}_n = \mathbf{K} \mathbf{f}_n + \mathbf{n}_n,
\end{equation}
for $n=1,\ldots,N$, 
where $\mathbf{m}_n \in \mathbb{R}^{ P}$ is the vector of $P$ different measured image voxel values $m_p(\mathbf{r}_n)$ from the $n$th voxel; $\mathbf{n}_n \in \mathbb{R}^{P}$ is the vector of noise values  $\eta_p(\mathbf{r}_n)$ from the $n$th voxel; $\mathbf{K} \in \mathbb{R}^{P\times Q}$ is the ``dictionary'' describing the ideal imaging physics, with entries $w_p b(\boldsymbol{\uptheta}_p,\boldsymbol{\upgamma}_q)$; and $\mathbf{f}_n\in \mathbb{R}^{Q}$ is the vector of the $Q$ spectral values $f(\boldsymbol{\upgamma}_q,\mathbf{r}_n)$ from the $n$th spatial location.  To simplify notation, we will also sometimes represent this expression as
\begin{equation}
 \mathbf{m} = (\mathbf{I}_N \otimes \mathbf{K}) \mathbf{f} + \mathbf{n},
\end{equation}
where $\mathbf{m} \in\mathbb{R}^{PN}$ is obtained by concatenating the vectors $\mathbf{m}_1,\ldots,\mathbf{m}_N$ into a long vector, and with similar concatenation operations applied to form the spectroscopic image vector $\mathbf{f} \in \mathbb{R}^{QN}$ and the noise vector $\mathbf{n} \in \mathbb{R}^{PN}$.  In this expression $\mathbf{I}_N$ represents the $N\times N$ identity matrix, and $\otimes$ represents the standard Kronecker product.

\subsection{Classical Voxel-By-Voxel Solutions}\label{sec:vbv}

Classically (e.g., \cite{Whittall1989}), the multicomponent mixture model is estimated by solving a regularized nonnegative least squares (NNLS) estimation problem independently for each voxel:
\begin{equation}
 \hat{\mathbf{f}}_n = \arg\min_{\substack{{\mathbf{f}_n} \in \mathbb{R}^Q\\ {\mathbf{f}_n} \geq \mathbf{0} }} \frac{1}{2}\| \mathbf{m}_n - \mathbf{K}\mathbf{f}_n\|_2^2 + \lambda R(\mathbf{f}_n),\label{eq:voxelbyvoxel}
\end{equation}
for $n=1,\ldots,N$.\footnote{Note that the use of least-squares in this expression implicitly corresponds to Gaussian noise-modeling assumptions, though this same least-squares formulation can be  easily adapted to Rician or noncentral chi noise statistics using half-quadratic majorize-minimize techniques \cite{varadarajan2015}.}  In Eq.~\eqref{eq:voxelbyvoxel},  the constraint $\mathbf{f}_n \geq \mathbf{0}$ should be interpreted elementwise and corresponds to a physically-motivated nonnegativity constraint on the spectroscopic image $f(\boldsymbol{\upgamma},\mathbf{r})$, while $R(\cdot)$ is an optional voxelwise regularization penalty and $\lambda$ is a regularization parameter.  Note that if $R(\cdot)$ is chosen to be a convex function, then Eq.~\eqref{eq:voxelbyvoxel} is a convex optimization problem and can be globally optimized using a variety of different convex optimization methods \cite{Boyd2004}. Interestingly, NNLS solutions often naturally produce very sparse solutions even without additional regularization \cite{Slawski2013}. 

One of the most commonly-used algorithms to solve unregularized or Tikhonov-regularized NNLS problems in this context is the active set algorithm by Lawson and Hanson\cite{Lawson1995}. Theoretically, this will converge to a globally optimal solution in a finite number of iterations, where the number of iterations generally scales with the number of unknowns \cite{Lawson1995}.  For voxel-by-voxel estimation (with a small number of unknowns), this algorithm  can be very fast and efficient, although the lack of spatial regularity constraints can lead to relatively poor results.  

\subsection{Spatially-Regularized Solutions}

In contrast to the voxel-by-voxel approach, the spatially-regularized approach estimates all voxels simultaneously in a coupled fashion, by solving a problem of the form
\begin{equation}
\begin{split}
 \hat{\mathbf{f}} = \arg\min_{\substack{\mathbf{f} \in \mathbb{R}^{QN} \\ \mathbf{f} \geq \mathbf{0}}} \frac{1}{2} \| \mathbf{m} - (\mathbf{I}_N \otimes \mathbf{K})\mathbf{f}\|_2^2 + \lambda R(\mathbf{f}),
 \end{split}\label{eq:spat}
\end{equation}
where the regularization penalty $R(\cdot)$ is chosen to encourage similarity between the spectra obtained at adjacent voxels.  For example, Refs.~\cite{Daeun2017,Daeun2020} made use of a Tikhonov regularization penalty\footnote{The choice to use Tikhonov regularization in Refs.~\cite{Daeun2017,Daeun2020} may appear unusual at first glance, as it has become common in recent years to use more advanced spatial regularizers (e.g., total variation). However, the choice of Tikhonov regularization was driven by the desire for the spatial regularization to behave gracefully given the ill-posed nature of the inverse problem \cite{Daeun2020}. While more advanced regularizers like total variation can preserve edges, they are also more likely to hallucinate nonexistent features when the problem is ill-posed.  On the other hand, Tikhonov regularization may produce undesirable blurring, but is much less likely to hallucinate.  At the same time, we are operating in contexts where we expect to see partial volume effects, where gradual transitions may be more likely to be observed than sharp edges. Note also that Tikhonov regularizers are convex, so that this choice also ensures that Eq.~\eqref{eq:spat} is convex and has a unique globally optimal solution. } of the form
\begin{equation}
\begin{split}
 R(\mathbf{f}) &= \sum_{n=1}^N \sum_{m \in \Delta(n)} \frac{1}{2} \| \mathbf{f}_n - \mathbf{f}_m\|_2^2 \triangleq \frac{1}{2} \|\mathbf{D}\mathbf{f}\|_2^2,
 \end{split}\label{eq:l2}
\end{equation}
where $\Delta(n)$ is the set of indices for all voxel that are adjacent to the $n$th voxel, and $\mathbf{D}$ is a matrix representation of the spatial finite difference operator. The remainder of this paper will assume this choice of $R(\cdot)$.

In principle, the Lawson-Hanson algorithm (see Sec.~\ref{sec:vbv}) could also be applied to solve the spatially-regularized problem from Eq.~\eqref{eq:spat}.  However, this requires increasing the number of unknowns by a factor of $N$ (the number of voxels), which leads to substantial increases in memory usage and  the number of iterations.  In practice, it has been observed that the Lawson-Hanson algorithm is only worthwhile to use when $N$ is quite small, which has motivated the use of modified problem formulations that only reconstruct small image patches instead of the entire image  \cite{Kumar2012, Labadie2014, Kumar2018}.  While this patch-based spatially-regularized approach can be somewhat computationally effective, the focus on image patches can introduce boundary artifacts at the edges of each patch that would not be observed had the entire image been reconstructed simultaneously.  

Several large-scale algorithms that accommodate large $N$ have also been proposed based on the ADMM algorithm \cite{Daeun2017,Daeun2020,Zimmermann2019}.  We will introduce the general mathematical principles of ADMM and LADMM for solving generic optimization problems in the sequel.  Below, we provide a quick description of how ADMM was adapted to solving Eq.~\eqref{eq:spat} in Ref.~\cite{Daeun2017}, which is representative of other implementations.  Ref.~\cite{Daeun2017} uses variable splitting to convert Eq.~\eqref{eq:spat} into an equivalent form:
\begin{equation}
\begin{split}
 \hat{\mathbf{f}} = \arg\min_{\mathbf{f}} \min_{\mathbf{x},\mathbf{y},\mathbf{z}} & \frac{1}{2} \|\mathbf{m} - (\mathbf{I}_N\otimes\mathbf{K})\mathbf{x}\|_2^2  + \mathcal{I}_+(\mathbf{y}) + \frac{\lambda}{2} \|\mathbf{D}\mathbf{z}\|_2^2,
 \end{split}\label{eq:const1}
\end{equation}
subject to the constraints that $\mathbf{f} = \mathbf{x}$, $\mathbf{f} = \mathbf{y}$, and $\mathbf{f} = \mathbf{z}$, where $\mathbf{x}$, $\mathbf{y}$ and $\mathbf{z}$ are new auxiliary variables that have been introduced to simplify subproblems within the  ADMM procedure.  In this expression, $\mathcal{I}_+(\cdot)$ is the indicator function for the feasible set of the non-negativity constraint \cite{afonso2011}:
\begin{equation}
 \mathcal{I}_+(\mathbf{z}) = \left\{ \begin{array}{ll} 0, & \mathbf{z} \geq 0 \text{ elementwise}, \\ +\infty, & \text{otherwise.} \end{array} \right.
\end{equation}
The algorithm then alternates between solving subproblems for $\mathbf{x}$, $\mathbf{y}$, and $\mathbf{z}$.  For completeness, the full ADMM implementation from Ref.~\cite{Daeun2017} is presented as Algorithm~\ref{alg:ADMM_Daeun}.\footnote{The original algorithm from Ref.~\cite{Daeun2017} involved masking of voxels outside the object. However, we have recently observed that we get very similar results (with better computational efficiency and a simpler algorithm description) if we do not use masks, but instead only estimate voxels inside  the object boundary without imposing spatial smoothness constraints across this boundary. As such, Algorithm~\ref{alg:ADMM_Daeun} is presented in a simplified maskless form.  } Note that this implementation requires storing at least 8 vectors that are the same size as the spectroscopic image $\mathbf{f}$, which can require substantial memory when $Q$ and $N$ are  large.  

\begin{algorithm}
	\caption{Previous ADMM Algorithm \cite{Daeun2017}}
	\label{alg:ADMM_Daeun}
	\begin{algorithmic}[1]
		\vspace{10pt}
		\STATE \textbf{Input:} ADMM parameter $\beta$. \\		
		\hspace{0.34in} Problem specification $\mathbf{K}$, $\mathbf{m}$, $\lambda$, $\mathbf{D}$.
		
		\hrulefill
		\STATE \textbf{Initialize:} Iteration number $k=0$.\\	\hspace{0.58in} Primal variables $\mathbf{f}^k$, $\mathbf{x}^k$, $\mathbf{y}^k$, $\mathbf{z}^k$  (we init to $\mathbf{0}$).	\\	\hspace{0.58in} Dual variables $\mathbf{d}_x^k$, $\mathbf{d}_y^k$, $\mathbf{d}_z^k$  (we init to $\mathbf{0}$).
\STATE $\mathbf{M} \triangleq \left( \mathbf{K}^T\mathbf{K} + \beta\mathbf{I}_Q \right)^{-1}$.
		\STATE $\mathbf{g} \triangleq  (\mathbf{I}_N \otimes \mathbf{K}^T) \mathbf{m}$.
 \\[5pt]
		\WHILE {not converged}
		\STATE $\mathbf{f}^{k+1} = \left( \beta \mathbf{x}^k + \mathbf{d}_x^k + \beta \mathbf{y}^k + \mathbf{d}_y^k + \beta \mathbf{z}^k + \mathbf{d}_z^k \right) / (3\beta) $\\[5pt]		
		\STATE $\mathbf{x}^{k+1} = \left(\mathbf{I}_N \otimes \mathbf{M}\right) \left(\mathbf{g}+ \beta \mathbf{f}^{k+1} - \mathbf{d}_x^{k}  \right)$ \\[5pt]		
		\STATE $\mathbf{y}^{k+1} = \max\left(\mathbf{0}, \mathbf{f}^{k+1} - \mathbf{d}_{y}^k/\beta \right)$ \\\hspace{0.5in}(maximization performed elementwise) \\[5pt]
		\STATE $\displaystyle\mathbf{z}^{k+1} = \left( \lambda\mathbf{D}^T\mathbf{D} + \beta\mathbf{I} \right)^{-1} \left( \beta \mathbf{f}^{k+1} - \mathbf{d}_{z}^k \right)$ \\[5pt]
		\STATE $\mathbf{d}_{x}^{k+1} = \mathbf{d}_{x}^k - \beta(\mathbf{f}^{k+1} - \mathbf{x}^{k+1})$ \\[5pt]
		\STATE $\mathbf{d}_{y}^{k+1} = \mathbf{d}_{y}^{k} - \beta(\mathbf{f}^{k+1} - \mathbf{y}^{k+1})$ \\[5pt]
		\STATE $\mathbf{d}_z^{k+1} = \mathbf{d}_{z}^{k} - \beta(\mathbf{f}^{k+1} - \mathbf{z}^{k+1})$ \\[5pt]
		\STATE $k \gets k + 1$
		\ENDWHILE
	\end{algorithmic}
\end{algorithm}

\section{Proposed LADMM-Based Approach}\label{sec:proposed}
\subsection{Review of ADMM and LADMM}\label{sec:LADMM}
Our proposed approach is based on LADMM \cite{He2002,Zhang2011,He2020,Fazel2013,Tao2020,Deng2016}, which itself is based on ADMM. We begin this section with a quick review. Both ADMM and LADMM are designed to solve generic optimization problems of the form 
\begin{equation}
 \label{eq:LADMM_prob}
\{\hat{\mathbf{p}},\hat{\mathbf{q}}\} = \arg\min_{\substack{\mathbf{p}\in \mathbb{R}^S \\\mathbf{q}\in\mathbb{R}^T}} \phi(\mathbf{p}) + \psi(\mathbf{q}) \quad s.t. \quad \mathbf{Ap + Bq = c},
\end{equation}
where $\phi(\cdot): \mathbb{R}^S \to \mathbb{R} \cup \{+\infty\}$ and $\psi(\cdot): \mathbb{R}^T \to \mathbb{R} \cup \{+\infty\}$  are closed proper convex functions. The corresponding augmented Lagrangian is
\begin{equation}
\begin{split}
	L_{\beta}(\mathbf{p,q,d}) &= \phi(\mathbf{p}) + \psi(\mathbf{q}) - \left\langle\mathbf{Ap + Bq - c}, \mathbf{d}\right\rangle \\ 
	&+ \frac{\beta}{2} \|\mathbf{Ap + Bq - c}\|_2^2,
\end{split}
\end{equation} 
where $\mathbf{d}$ is the dual vector and $\beta > 0$ is a penalty parameter. 

Conventional ADMM \cite{Boyd2011} would approach the optimization problem from Eq.~\eqref{eq:LADMM_prob} by alternatingly updating the primal variables $\mathbf{p}$, $\mathbf{q}$ and the dual vector $\mathbf{d}$ according to
\begin{equation} \label{eq:ADMM_alg}
\begin{split}
\begin{cases}
	\displaystyle\mathbf{q}^{k+1} = \arg\min_{\mathbf{q}} L_{\beta}(\mathbf{p}^{k}, \mathbf{q}, \mathbf{d}^k) \\\displaystyle
	\mathbf{p}^{k+1} = \arg\min_{\mathbf{p}} L_{\beta}(\mathbf{p}, \mathbf{q}^{k+1}, \mathbf{d}^k) \\
	\mathbf{d}^{k+1} = \mathbf{d}^k - \beta(\mathbf{Ap}^{k+1} + \mathbf{Bq}^{k+1} - \mathbf{c}).
\end{cases}
\end{split}
\end{equation}
By completing the square, the $\mathbf{p}$ and $\mathbf{q}$ subproblems  of Eq.~\eqref{eq:ADMM_alg} can be further simplified to
\begin{equation}
\begin{split}
\begin{cases}
	\displaystyle\mathbf{q}^{k+1} = \arg\min_{\mathbf{q}} \psi(\mathbf{q}) + \frac{\beta}{2}\left\| \mathbf{B}\mathbf{q} - \left(\mathbf{c} - \mathbf{A}\mathbf{p}^{k} + \mathbf{d}^k / \beta \right)\right\|_2^2 \\\displaystyle
	\mathbf{p}^{k+1} = \arg\min_{\mathbf{p}} \phi(\mathbf{p}) + \frac{\beta}{2}\left\| \mathbf{A}\mathbf{p} - \left(\mathbf{c} - \mathbf{B}\mathbf{q}^{k+1} + \mathbf{d}^k / \beta \right)\right\|_2^2.
\end{cases}
\end{split}
 \label{eq:ADMM_sub}
\end{equation}
Indeed, the previous ADMM approach (Algorithm 1) can be obtained by making the associations
\begin{equation}
	\begin{split}
		&\mathbf{p} \in \mathbb{R}^S \rightarrow \begin{bmatrix}  \mathbf{x}^T & \mathbf{y}^T & \mathbf{z}^T	\end{bmatrix}^T \in \mathbb{R}^{3QN}, \\
&  \mathbf{q} \in \mathbb{R}^T \rightarrow \mathbf{f} \in \mathbb{R}^{QN},\:\:\:\:\:\:\:\:\:\:\:\:\:\:\:\:\:\:\mathbf{c} \rightarrow \mathbf{0}\in\mathbb{R}^{3QN}, \\
		&\mathbf{A} \rightarrow \mathbf{I}_{3QN}, \:\:\:\:\:\:\:\:\:\:\:\:\:\:\:\:\:\: \mathbf{B} \rightarrow -\begin{bmatrix}  \mathbf{I}_{QN} & \mathbf{I}_{QN} & \mathbf{I}_{QN}	\end{bmatrix}^T,  \\
		&\phi(\mathbf{p}) \rightarrow \frac{1}{2} \|\mathbf{m} - (\mathbf{I}_N\otimes\mathbf{K})\mathbf{x}\|_2^2 + \mathcal{I}_+(\mathbf{y}) + \frac{\lambda}{2} \|\mathbf{D}\mathbf{z}\|_2^2, \\
		&\psi(\mathbf{q}) \rightarrow 0.
	\end{split}
\end{equation}
The speed of ADMM depends on the ease of solving the subproblems in Eq.~\eqref{eq:ADMM_sub} and how quickly the iterations converge. 

LADMM builds upon ADMM, but introduces additional quadratic terms involving $\mathbf{P}$ and $\mathbf{Q}$ matrices which can be chosen to simplify the solution of subproblems. Specifically, LADMM iteratively updates variables according to
\begin{equation} \label{eq:LADMM_alg}
\begin{cases}\displaystyle
	\mathbf{q}^{k+1} = \arg\min_{\mathbf{q}} L_{\beta}(\mathbf{p}^{k}, \mathbf{q}, \mathbf{d}^k) + \frac{1}{2}(\mathbf{q - q}^k)^T \mathbf{Q} (\mathbf{q - q}^k) \\\displaystyle
	\mathbf{p}^{k+1} = \arg\min_{\mathbf{p}} L_{\beta}(\mathbf{p}, \mathbf{q}^{k+1}, \mathbf{d}^k) + \frac{1}{2}(\mathbf{p - p}^k)^T \mathbf{P} (\mathbf{p - p}^k) \\[7pt]
	\mathbf{d}^{k+1} = \mathbf{d}^k - \beta(\mathbf{Ap}^{k+1} + \mathbf{Bq}^{k+1} - \mathbf{c}).
\end{cases}
\end{equation} 
Detailed discussion of choosing  $\mathbf{P}$ and $\mathbf{Q}$ matrices is found in Ref.~\cite{Deng2016}. One of the standard choices  \cite{He2002,Zhang2011,He2020,Fazel2013,Tao2020,Deng2016} is to use
\begin{equation} \label{eq:LADMM_PQ}
\begin{split}
	\mathbf{P} = \xi_p \mathbf{I}_S - \beta \mathbf{A}^T\mathbf{A} \text{ and } 	\mathbf{Q} = \xi_q \mathbf{I}_T - \beta \mathbf{B}^T\mathbf{B} 
\end{split},
\end{equation} 
with proper choices of parameters $\xi_p$ and $\xi_q$. Importantly, these choices cause the $\mathbf{p}$ and $\mathbf{q}$ subproblems from Eq.~\eqref{eq:LADMM_alg} to simplify substantially, in a way that cancels some of the $\mathbf{A}$ and $\mathbf{B}$ matrices.  Specifically, by again completing the square, the $\mathbf{p}$ and $\mathbf{q}$ subproblems from Eq.~\eqref{eq:LADMM_alg} reduce to 
\begin{equation} \label{eq:LADMM_alg_ex}
\begin{split}
\begin{cases}
	\displaystyle \mathbf{q}^{k+1} = \arg\min_{\mathbf{q}} \psi(\mathbf{q}) + \frac{\xi_p}{2} \|\mathbf{q} - \tilde{\mathbf{q}}^k \|_2^2 \\
	\displaystyle \mathbf{p}^{k+1} = \arg\min_{\mathbf{p}} \phi(\mathbf{p}) + \frac{\xi_q}{2} \|\mathbf{p} - \tilde{\mathbf{p}}^k \|_2^2, 
\end{cases}
\end{split}
\end{equation} 
where 
\begin{equation}
\begin{split}
	\tilde{\mathbf{q}}^k &= \mathbf{q}^k - (\beta / \xi_q) \mathbf{B}^T (\mathbf{Ap}^k + \mathbf{Bq}^k - \mathbf{c - d}^k / \beta) \\
	\tilde{\mathbf{p}}^k &= \mathbf{p}^k - (\beta / \xi_p) \mathbf{A}^T (\mathbf{Ap}^k + \mathbf{Bq}^{k+1} - \mathbf{c - d}_k / \beta).
\end{split}
\end{equation}
The two subproblems in Eq.~\eqref{eq:LADMM_alg_ex} are equivalent to evaluating the proximal operators of the functions $\phi(\cdot)$ and $\psi(\cdot)$.  This is beneficial because proximal operators are known in closed-form for a variety of interesting functions, including  the $\ell_1$-norm and the squared $\ell_2$-norm  \cite{Combettes2011,Boyd2014,Beck2017}, which enables simple/fast solution of the subproblems.  In our proposed approach, presented in the sequel, we will make different choices $\mathbf{P}$ and $\mathbf{Q}$ than those given in Eq.~\eqref{eq:LADMM_PQ}, although our choice of $\mathbf{P}$ is motivated by Eq.~\eqref{eq:LADMM_PQ} and the preceding discussion.

Convergence of LADMM has been analyzed for different scenarios.  In \cite{He2002,Zhang2011}, it was proven that LADMM will converge to a globally-optimal solution from an arbitrary initialization whenever $\mathbf{P}$ and $\mathbf{Q}$ are symmetric and positive definite.  In \cite{Fazel2013}, it was shown that the same kind of global convergence will be obtained whenever $\mathbf{P}$ and $\mathbf{Q}$ are symmetric and positive semi-definite.  In \cite{He2020}, it was shown that the same kind of global convergence will also be obtained if $\mathbf{P}$ and $\mathbf{Q}$ are chosen as in Eq.~\eqref{eq:LADMM_PQ}, but with $\xi_p > 0.75\|\mathbf{A}^T\mathbf{A}\|$ and $\xi_q > 0.75\|\mathbf{B}^T\mathbf{B}\|$ (which enables the use of certain indefinite $\mathbf{P}$ and $\mathbf{Q}$ matrices), where $\|\cdot\|$ denotes the spectral norm.  Several papers have also derived even less-restrictive conditions on $\mathbf{P}$ and $\mathbf{Q}$ that guarantee global convergence \cite{He2020,Tao2020,Deng2016}.  The algorithm we propose in the sequel makes use of a positive semi-definite $\mathbf{P}$ matrix and an indefinite $\mathbf{Q}$ matrix that satisfies the constraints from \cite{He2020}, and thus inherits the global convergence guarantees from previous literature.

\subsection{Proposed Method}\label{sec:propm}
Our proposed approach is based on applying LADMM principles to the optimization problem from Eq.~\eqref{eq:spat}. Instead of using three variable splittings as was done in Eq.~\eqref{eq:const1} for  ADMM  \cite{Daeun2017}, we propose to use a formulation based on a single variable splitting that is also equivalent to Eq.~\eqref{eq:spat}:
 \begin{equation} \label{eq:Optprob_const}
 \begin{split}
\hat{\mathbf{f}}	= \arg\min_\mathbf{f} \min_{\mathbf{z}} & \frac{1}{2} \|\mathbf{m} - (\mathbf{I}_N \otimes \mathbf{K)f}\|_2^2  + \mathcal{I}_+(\mathbf{z}) 
 + \frac{\lambda}{2} \|\mathbf{D}\mathbf{z}\|_2^2, 
	\end{split}
\end{equation}
subject to the constraint that $\mathbf{f} = \mathbf{z}$.  With this, we can map  the LADMM problem described in the previous subsection to our desired optimization problem by making the associations:
\begin{equation}
 \begin{split}
  &\mathbf{p} \in \mathbb{R}^S \rightarrow \mathbf{z} \in \mathbb{R}^{QN},\:\:\:\:\:\:\:\:\:\:\:\:\:\:   \mathbf{q} \in \mathbb{R}^T \rightarrow \mathbf{f} \in \mathbb{R}^{QN}  \\
  &\mathbf{c} \rightarrow \mathbf{0}\in\mathbb{R}^{QN}, \:\:\:\:\:\:\:\:\:\:\:\:\:\: \mathbf{A} \rightarrow \mathbf{I}_N, \:\:\:\:\:\:\:\:\:\:\:\:\:\:  \mathbf{B} \rightarrow -\mathbf{I}_N, \\
  &\phi(\mathbf{p}) \rightarrow \mathcal{I}_+(\mathbf{x}) + \frac{\lambda}{2} \|\mathbf{D}\mathbf{x}\|_2^2,  \\
  &\psi(\mathbf{y}) \rightarrow \frac{1}{2} \|\mathbf{m} - (\mathbf{I}_N\otimes \mathbf{K}) \mathbf{y}\|_2^2.
 \end{split}
\end{equation}

This directly results in the  set of LADMM updates:
\begin{equation}
\begin{split}
 \mathbf{f}^{k+1} &= \arg\min_{\mathbf{f}} \frac{1}{2} \|\mathbf{m} - (\mathbf{I}_N \otimes \mathbf{K})\mathbf{f}\|_2^2 - \left\langle \mathbf{z}^k-\mathbf{f},\mathbf{d}^k\right\rangle \\
 \,&\hspace{0.3in}+ \frac{\beta}{2} \left\|\mathbf{z}^k - \mathbf{f}  
 \right\|_2^2 + \frac{1}{2} ( \mathbf{f} - \mathbf{f}^k)^T \mathbf{Q} ( \mathbf{f} - \mathbf{f}^k),
 \end{split}\label{eq:f}
\end{equation}
\begin{equation}
\begin{split}
 \mathbf{z}^{k+1} &= \arg\min_{\mathbf{z}} \mathcal{I}_+(\mathbf{z})  + \frac{\lambda}{2} \|\mathbf{D}\mathbf{z}\|_2^2 - \left\langle \mathbf{z}-\mathbf{f}^{k+1},\mathbf{d}^k\right\rangle \\
 \,&\hspace{0.3in}+  \frac{\beta}{2} \| \mathbf{z} - \mathbf{f}^{k+1}\|_2^2+ \frac{1}{2} ( \mathbf{z} - \mathbf{z}^k)^T \mathbf{P} ( \mathbf{z} - \mathbf{z}^k),
 \end{split}\label{eq:z}
\end{equation}
and finally
\begin{equation}
 \mathbf{d}^{k+1} = \mathbf{d}^k - \beta (\mathbf{z}^{k+1} - \mathbf{f}^{k+1}).
\end{equation}

Our choices of $\mathbf{P}$ and $\mathbf{Q}$ and the methods we use to solve these subproblems are based on novel insights into the structure of the optimization problem, as described in the following subsections.

\subsection{Solving the $\mathbf{f}$ subproblem}
It is easy to show that  Eq.~\eqref{eq:f} has the same solution as
\begin{equation}
\begin{split}
 \hspace{-0.05in} \mathbf{f}^{k+1} &= \arg\min_{\mathbf{f}} \frac{1}{2} \|\mathbf{m} - (\mathbf{I}_N \otimes \mathbf{K})\mathbf{f}\|_2^2 + \frac{\beta}{2} \left\|\mathbf{f} - \mathbf{z}^k +\mathbf{d}^k/\beta 
 \right\|_2^2 \\
 \,&\hspace{0.6in}+  \frac{1}{2} ( \mathbf{f} - \mathbf{f}^k)^T \mathbf{Q} ( \mathbf{f} - \mathbf{f}^k),
 \end{split}\label{eq:f2}
\end{equation}
Notably, if we choose $\mathbf{Q = 0}$ (which is positive semi-definite), then the resulting problem ends up having a decoupled (separable) block-diagonal structure, which allows the solution to be computed in a simple voxel-by-voxel manner.  Compared to the voxel-by-voxel NNLS reconstruction problem from Eq.~\eqref{eq:voxelbyvoxel}, the $\mathbf{f}$ subproblem is even simpler because it does not involve any nonnegativity constraints, which allows the optimal solution to be computed analytically as:
\begin{equation} \label{eq:DictMatxInv}
 \mathbf{f}^{k+1}_n = \mathbf{M} \left(\mathbf{K}^T\mathbf{m}_n + \beta \mathbf{z}_n^k - \mathbf{d}_n^k\right),
\end{equation}
where $\mathbf{z}_n^k \in \mathbb{R}^Q$ and $\mathbf{d}_n^k\in \mathbb{R}^Q$ are respectively the components of $\mathbf{z}^k$ and $\mathbf{d}^k$ corresponding to the $n$th voxel position, and  the matrix $\mathbf{M} \in \mathbb{R}^{Q\times Q} \triangleq \left(\mathbf{K}^T\mathbf{K} + \beta \mathbf{I}_Q\right)^{-1}.$

Note that the matrix $\mathbf{M}$ is the same  for all voxels $n$ and all iterations $k$, which allows it to be precomputed once and then reused whenever it is needed \cite{Daeun2017}.  This approach is viable and efficient when $Q$ is small.  However, in scenarios where $Q$ is large (e.g., in the multidimensional multiparametric setting \cite{Daeun2017,Daeun2020,McGivney2018,deshmane2019,Tang2018,nagtegaal2020,slator2021,benjamini2020}), storing and performing matrix-vector multiplications with a $Q\times Q$ matrix can still be relatively computationally burdensome, requiring $O(Q^2)$ memory for storage and $O(Q^2)$ floating point operations for matrix-vector multiplication.

The computational complexity associated with $\mathbf{M}$ can be substantially reduced by leveraging the fact that while the matrix $\mathbf{K}^T\mathbf{K}$ is of size $Q \times Q$, it typically has rank that is smaller than $Q$.\footnote{Using low-rank structure to enable faster computations is common in numerical linear algebra. The ability to do this is not unique to LADMM, although the previous ADMM algorithms for this problem context did not exploit the low-rank structure.  }  Specifically, the matrix will have rank that is no greater than $P$, and it is frequently the case that $P < Q$.  In practice, it is also common that  $\mathbf{K}$ has highly-correlated columns  (approximately linearly dependent), allowing accurate approximation by a rank-$r$ matrix  with $r < P$ \cite{venkataramanan2002,Yang2018}.   

Let $\mathbf{K}_r \in \mathbb{R}^{P\times Q}$ be a rank-$r$ approximation of $\mathbf{K}$ obtained by truncating the singular valued decomposition (SVD) of $\mathbf{K}$:
\begin{equation} \label{eq:DictSVD}
 \mathbf{K}_r = \sum_{i=1}^r \sigma_i \mathbf{u}_i \mathbf{v}_i^T ,
\end{equation}
with singular vectors $\mathbf{u}_i \in \mathbb{R}^{P}$ and $\mathbf{v}_i\in \mathbb{R}^{Q}$ and singular values $\sigma_i>0$ for $i=1,\ldots,r$. It is straightforward to derive that matrix-vector multiplications involving the matrix $\mathbf{M}_r \triangleq \left(\mathbf{K}_r^T\mathbf{K}_r + \beta \mathbf{I}_Q\right)^{-1}$ associated with the rank-$r$ approximation of the dictionary can be calculated  simply as
\begin{equation} \label{eq:DictVecSVD}
 \mathbf{M}_r \mathbf{x} = \frac{1}{\beta} \mathbf{x} - \sum_{i=1}^r  \left(\frac{\sigma_i^2}{\beta^2 + \beta\sigma_i^2}\right) \mathbf{v}_i \left( \mathbf{v}_i^T\mathbf{x}\right)
\end{equation}
for arbitrary vectors $\mathbf{x} \in \mathbb{R}^Q$.  With this approximation, storing $\mathbf{M}_r$ requires only $O(rQ)$ memory (for storing the vectors $\mathbf{v}_i$ and singular values $\sigma_i$) and $O(rQ)$ floating point operations for computing matrix-vector multiplication, which can be a substantial improvement over the naive $O(Q^2)$ complexity associated with directly using the matrix $\mathbf{M}$.

\subsection{Solving the $\mathbf{z}$ subproblem}\label{sec:z}
It can also be shown that Eq.~\eqref{eq:z} has the same solution as 
\begin{equation}
 \begin{split}
  \mathbf{z}^{k+1} &= \arg\min_\mathbf{z} \mathcal{I}_+(\mathbf{z}) + \frac{\beta}{2} \| \mathbf{z} -\mathbf{f}^{k+1} -  \mathbf{d}^k/\beta\|_2^2 \\
  \,&\hspace{0.6in}+ \frac{\lambda}{2} \|\mathbf{D}\mathbf{z}\|_2^2 + \frac{1}{2} ( \mathbf{z} - \mathbf{z}^k)^T \mathbf{P} ( \mathbf{z} - \mathbf{z}^k).
 \end{split}\label{eq:z3}
\end{equation}
In this case, the spatial finite difference operator $\mathbf{D}$ causes coupling of the different spatial locations, which makes the nonnegativity-constrained problem difficult to solve analytically.  In previous ADMM approaches \cite{Daeun2017, Daeun2020, Zimmermann2019}, this issue is addressed by introducing more splitting variables, which can increase memory consumption and reduce convergence speed.

In our proposed approach, motivated by Eq.~\eqref{eq:LADMM_PQ} and the associated discussion from Sec.~\ref{sec:LADMM}, we choose the matrix $\mathbf{P}$ to simplify the problem so that we can solve for $\mathbf{z}^{k+1}$  directly.  Specifically, we choose 
\begin{equation}
 \mathbf{P} = \xi_p \mathbf{I}_{QN} - \lambda \mathbf{D}^T\mathbf{D},
\end{equation}
with some parameter $\xi_p$ to be determined later.  This choice leads to cancellation of $\mathbf{D}$, fully decoupling the optimization.

With this choice,  if we ignore the nonnegativity constraint on the elements of $\mathbf{z}^{k+1}$, the optimal solution for the $\mathbf{z}$ subproblem becomes
\begin{equation}
 \mathbf{z}^{k+1} = \frac{1}{\xi_p + \beta}\left(\xi_p \mathbf{z}^k - \lambda \mathbf{D}^T\mathbf{D} \mathbf{z}^k + \beta \mathbf{f}^{k+1} + \mathbf{d}^{k}\right),\label{eq:z2}
\end{equation}
which can be computed easily and  noniteratively.  Because of the simple decoupled structure, imposing the nonnegativity constraint is as simple as taking the positive entries of $\mathbf{z}^{k+1}$ from Eq.~\eqref{eq:z2},  replacing all of the negative entries  with 0.

\subsection{Proposed Algorithm}

The  proposed algorithm is presented as Algorithm~\ref{alg:LADMM4CSI}.

\begin{algorithm}
	\caption{Proposed LADMM Algorithm}
	\label{alg:LADMM4CSI}
	\begin{algorithmic}[1]
		\vspace{10pt}
		\STATE \textbf{Input:} LADMM parameters $\beta$, $\xi_p$, $r$. \\		
		\hspace{0.34in} Problem specification $\mathbf{K}$, $\mathbf{m}$, $\lambda$, $\mathbf{D}$.
		
		\hrulefill
		\STATE \textbf{Initialize:} Iteration number $k=0$.\\	\hspace{0.58in} Primal variables $\mathbf{f}^k$, $\mathbf{z}^k$ (we init to $\mathbf{0}$). 	\\	\hspace{0.58in} Dual variables $\mathbf{d}^k$ (we init to $\mathbf{0}$). 
		\STATE Compute $\mathbf{u}_i$, $\mathbf{v}_i$, and $\sigma_i$ from the SVD of $\mathbf{K}$. \\
        \STATE $\mathbf{g}_n \triangleq  \sum_{i=1}^r \sigma_i \mathbf{v}_i \mathbf{u}_i^T \mathbf{m}_n$, for $n=1,2,...,N$. \\[5pt]
		\WHILE {not converged}
		 \FOR{$n=1,2,...,N$} 
		\STATE $\mathbf{c}_n^k = \mathbf{g}_n + \beta \mathbf{z}_n^k - \mathbf{d}_n^k$
		\STATE $\mathbf{f}_n^{k+1} = \frac{1}{\beta} \mathbf{c}_n^k - \sum_{i=1}^r  \left(\frac{\sigma_i^2}{\beta^2 + \beta\sigma_i^2}\right) \mathbf{v}_i \left( \mathbf{v}_i^T\mathbf{c}_n^k \right)$
		\ENDFOR
		\STATE $\mathbf{z}_{k+1} = \max\left(\mathbf{0}, \frac{\xi_p \mathbf{z}^k - \lambda \mathbf{D}^T\mathbf{D} \mathbf{z}^k + \beta \mathbf{f}^{k+1} + \mathbf{d}^{k}}{\xi_p + \beta} \right)$ \\[5pt]
		\STATE $\mathbf{d}^{k+1} = \mathbf{d}^k - \beta (\mathbf{z}^{k+1} - \mathbf{f}^{k+1})$ \\[5pt]	
		\STATE $k \gets k + 1$
		\ENDWHILE
	\end{algorithmic}
\end{algorithm}

Given the choices described above, our proposed LADMM approach uses substantially fewer variables and has substantially fewer subproblems than the previous ADMM approach (Algorithm~\ref{alg:ADMM_Daeun}).  In addition, our implementation benefits from more efficient matrix inversion for the $\mathbf{f}$ subproblem based on the low-rank properties of $\mathbf{K}$, and eliminates the need for matrix inversion or additional splitting for the $\mathbf{z}$ subproblem.  

\subsection{Practical Parameter Selection}
Use of the proposed algorithm requires choosing the values of the parameters $\xi_p$, $\beta$, and $r$.
\subsubsection{The parameter $\xi_p$}
As described previously, convergence of Algorithm \ref{alg:LADMM4CSI} is guaranteed if $\xi_p$ is chosen such that $\xi_p > 0.75 \lambda \|\mathbf{D}^T\mathbf{D}\|$ \cite{He2020}. The results shown later simply use 
\begin{equation} \label{eq:sigma_cond}
	\xi_p = 0.75 \lambda \|\mathbf{D}^T\mathbf{D}\| + \epsilon,
\end{equation} 
where $\epsilon$ is a small positive number (we use $\epsilon=10^{-10}$).

\subsubsection{The penalty parameter $\beta$}
Although LADMM will converge to a globally optimal solution regardless of the choice of $\beta$, the choice of $\beta$ still has a big impact on the convergence speed \cite{Boyd2011}. Unfortunately, optimal methods for selecting $\beta$ are only known for a few  special cases \cite{Ghadimi2015}. Some adaptive methods for selecting $\beta$ have been proposed \cite{He2000,Xu2017}, although require additional computations that are not attractive for the type of large-scale problem considered in this work. In results shown later, we use a simple heuristic to select $\beta$ in a context-dependent yet computationally-efficient  way.  Specifically, we first consider the problem of estimating a small image patch (e.g., $3 \times 3$ voxels), and tune $\beta$ to maximize convergence speed for the small-scale problem.  We then use this same value to solve the full-scale optimization problem.

\subsubsection{The rank parameter $r$}
The choice of the rank parameter $r$ represents a trade-off between computational efficiency and the accuracy of the dictionary approximation.  For the examples shown in the sequel, the ranks were chosen to be the smallest values that produced less than $0.005\%$ approximation error (as measured in the Frobenius norm).

\begin{figure*}[t]
\begin{center}
\includegraphics[scale=1]{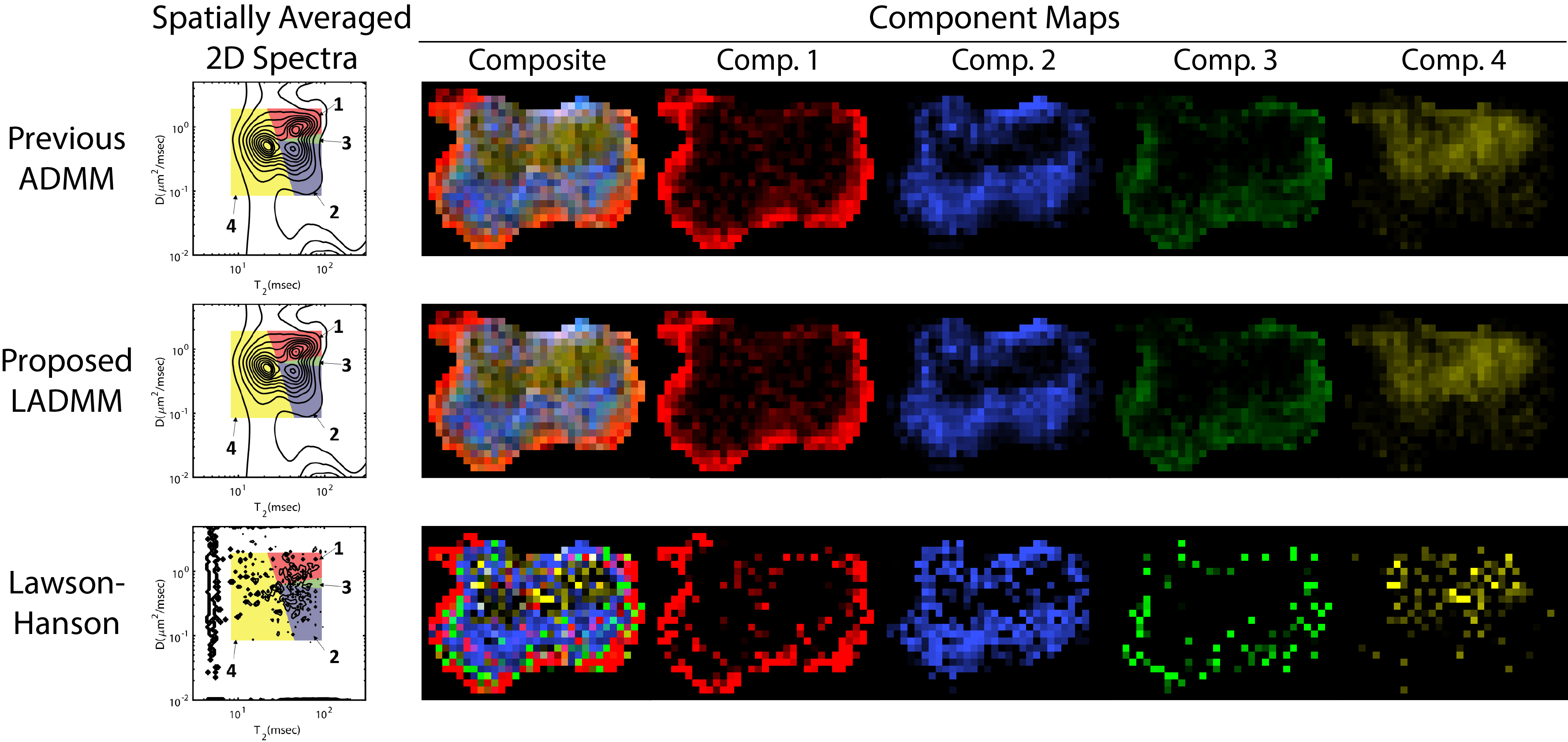}
\end{center}

	\caption{Component maps and spatially-averaged diffusion-relaxation spectra obtained for the DR-CSI data using ADMM, LADMM, and the voxel-by-voxel approach using the Lawson-Hanson algorithm.  The component maps were obtained by spectrally-integrating different regions (marked in the averaged spectra) of the 2D diffusion-relaxation spectrum \cite{Daeun2017,Daeun2020}. }
	\label{fig:DRCSI_comp}
\end{figure*}

\begin{figure}[t]
	\centering
	\subfloat{\includegraphics[width=1.6in]{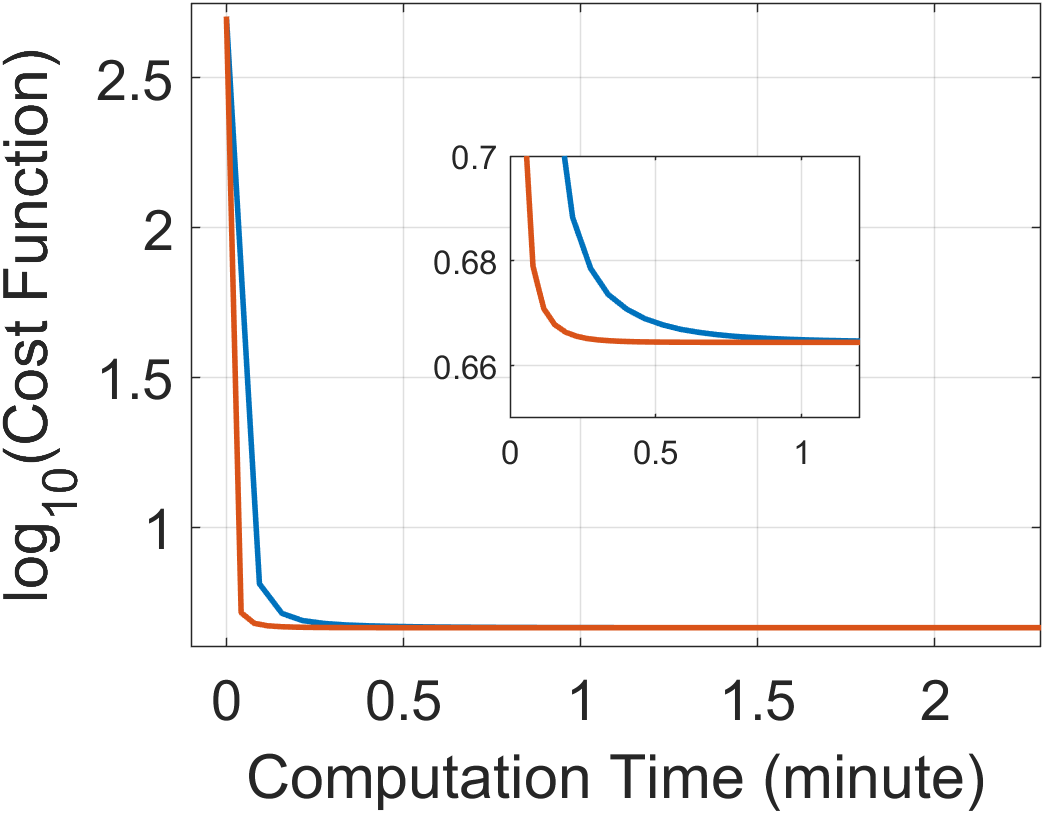}}
	\hfil
	\subfloat{\includegraphics[width=1.6in]{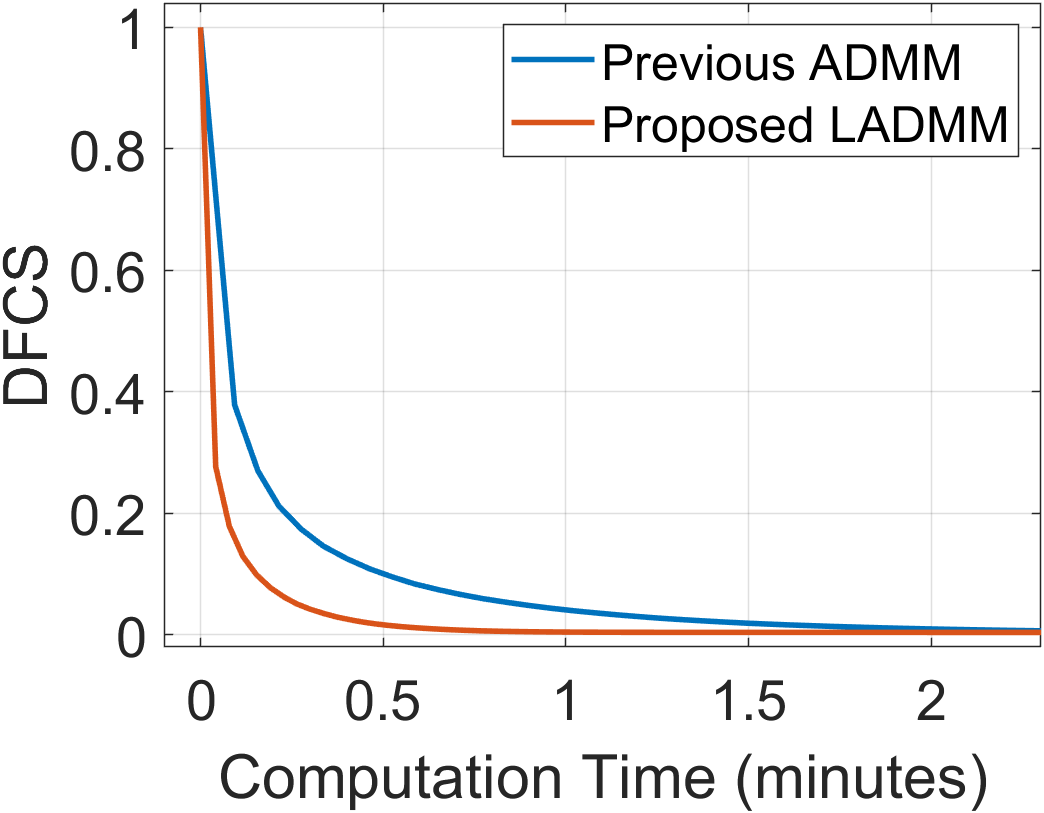}}
	\caption{Convergence of ADMM and LADMM for DR-CSI. }
	\label{fig:DRCSI_conv}
\end{figure}

\section{Results}
\label{sec:results}
We compared our new LADMM algorithm against the previous ADMM algorithm \cite{Daeun2017} in a range of different scenarios. To evaluate convergence for both ADMM and LADMM, we tracked the cost function value and the distance from the converged solution (DFCS) as a function of computation time, with DFCS defined as
\begin{equation} \label{eq:DFCS}
	\text{DFCS}(\mathbf{f}^k) \triangleq \frac{\|\mathbf{f}^k - \mathbf{f}^*\|_2}{\|\mathbf{f}^* \|_2}
\end{equation}
where $\mathbf{f}^k$ is  the estimated spectroscopic image at iteration $k$ and $\mathbf{f}^*$ is the result obtained after final convergence of a version of LADMM that was implemented using the original unapproximated $\mathbf{K}$ matrix.\footnote{Note that DFCS can be a better indicator of convergence than the cost function value, since the ill-posedness of the problem means that many different solutions will have similar cost function values.}  Note that the cost function value for $\mathbf{f}^*$ was always at least incrementally better than the best cost function values achieved by ADMM or LADMM with approximation, and therefore represents a good comparison reference.  All methods were implemented  in MATLAB on a server with an Intel Xeon E5-2690 2.6GHz 28-core CPU and 256 GB RAM.

The following subsections report results from four datasets from a range of different application scenarios. 

\subsection{DR-CSI }
We  evaluated LADMM on one of the diffusion-$T_2$ DR-CSI datasets from Ref.~\cite{Daeun2017}, corresponding to an ex vivo  injured mouse spinal cord     (specifically, the dataset described in Ref.~\cite{Daeun2017} as ``injured subject 1").  The details of this dataset and problem formulation were described in detail in Ref.~\cite{Daeun2017}, but some of the key parameters of the optimization problem were that data acquisition involved $P=28$ diffusion-relaxation encodings, the dictionary had $Q=4900$ elements, and the 2D image had $N=536$ voxels.  The resulting 4D spectroscopic image was comprised of a 2D diffusion-$T_2$ spectrum of size 70$\times$70 at each spatial image location. 

Fig.~\ref{fig:DRCSI_comp} shows  component maps and spatially-averaged diffusion-relaxation spectra obtained by both ADMM and LADMM in this case, while Fig.~\ref{fig:DRCSI_conv} compares their convergence characteristics.  As expected, the ADMM and LADMM component maps closely matched one another (and also match the spectra and component maps from Ref.~\cite{Daeun2017}, which contains a more detailed explanation and interpretation of these maps), demonstrating that the two algorithms yield similar results  if we allow them both enough time to sufficiently  (if we had stopped the computation at a fixed time prior to convergence, we would have seen more substantial differences as indicated by Fig.~\ref{fig:DRCSI_conv}).   Notably, LADMM converges much faster than ADMM (while this is true for both the cost function value and the DFCS, the cost function values in Fig.~\ref{fig:DRCSI_conv} appear to converge much faster than the DFCS -- this is not surprising given the ill-posed nature of the inverse problem, and our subsequent analysis will focus on DFCS).  For example, the DFCS value obtained by ADMM after 1.5 minutes of computation was achieved by LADMM in only 0.48 minutes (roughly a 3.1$\times$ improvement). LADMM also used substantially less memory than ADMM. Peak memory usage was 2.5 MB for LADMM versus 22.9 MB for ADMM (roughly a 9.1$\times$ improvement).

For reference, Fig.~\ref{fig:DRCSI_comp} also shows an example of the results obtained from voxel-by-voxel spectrum estimation using the Lawson-Hanson algorithm (see Sec.~\ref{sec:vbv}).  Although the voxel-by-voxel estimation results can be obtained very quickly (roughly 15 seconds of computation time) relative to spatially-regularized reconstruction, it is clear that the resulting spectra and spatial component maps qualitatively appear to be quite noisy, with less spectral coherence and less spatial correspondence with known anatomy.

\subsection{Inversion-Recovery Multi-Echo Spin-Echo RR-CSI}
We  also evaluated LADMM on one of the IR-MSE $T_1$-$T_2$  RR-CSI datasets from Ref.~\cite{Daeun2020}  of the human brain    (specifically, the dataset described in Ref.~\cite{Daeun2020} as ``subject 1").  The details of this dataset and problem formulation were described in detail in Ref.~\cite{Daeun2020}, but some of the key parameters of the optimization problem were that data acquisition involved $P=105$ $T_1$-$T_2$ relaxation encodings, the dictionary had $Q=10000$ elements, and the 2D image had $N=4553$ voxels. The resulting 4D spectroscopic image was comprised of a 2D $T_2$-$T_2$ spectrum of size 100$\times$100 at each spatial image location. 

\begin{figure*}[t]
	\begin{center}
\includegraphics[scale=1]{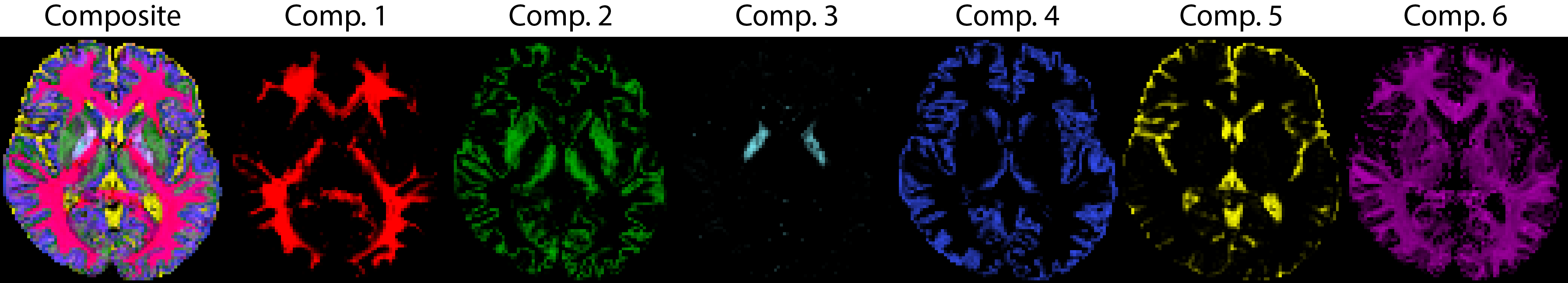}
	\end{center}
	\caption{Component maps obtained for the RR-CSI data using LADMM.  The component maps were obtained by spectrally-integrating different regions of the 2D $T_1$-$T_2$ spectrum as described in Ref.~\cite{Daeun2020}.}
	\label{fig:IRMSE_comp}
\end{figure*}

\begin{figure}[t]
	\centering
	\subfloat{\includegraphics[width=1.6in]{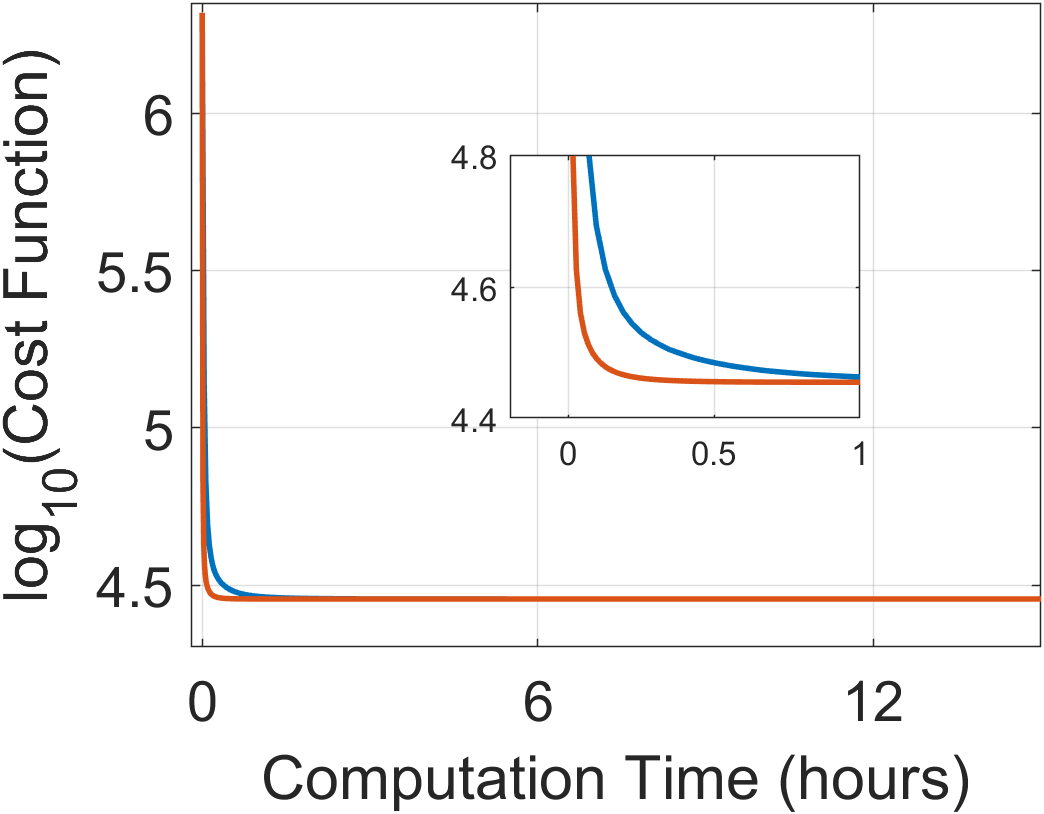}}
	\hfil
	\subfloat{\includegraphics[width=1.6in]{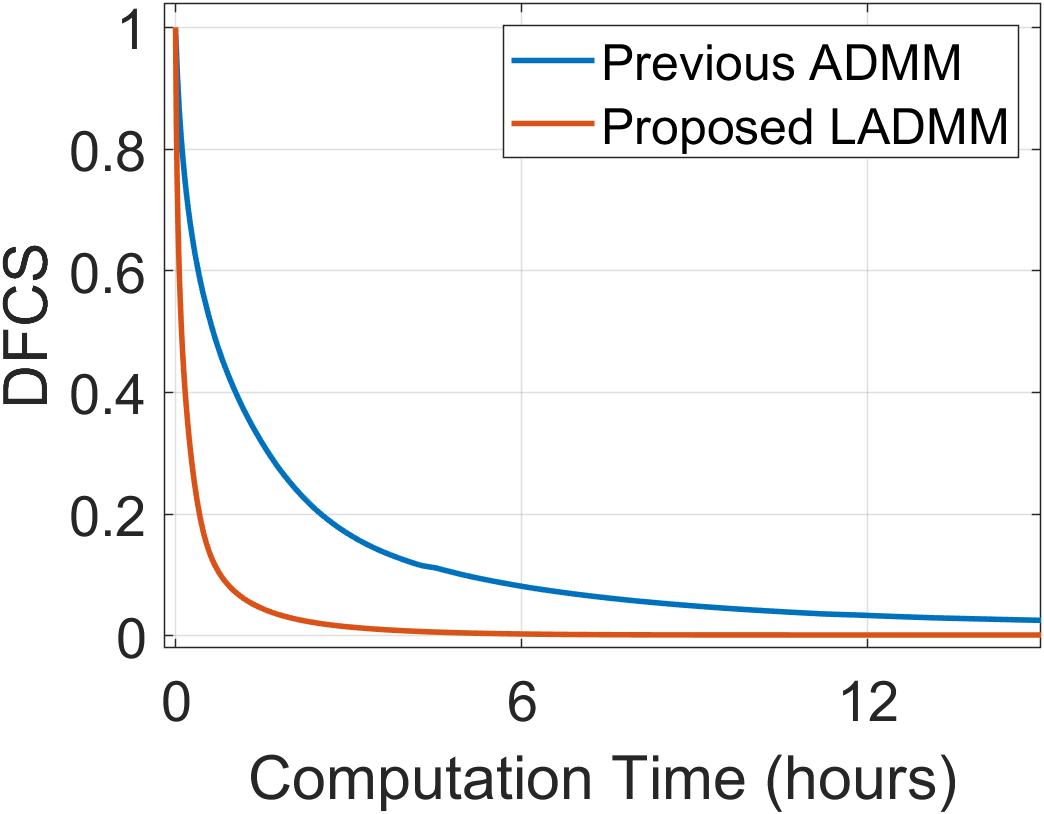}}
	\caption{Convergence of ADMM and LADMM for RR-CSI. }
	\label{fig:IRMSE_conv}
\end{figure}

Fig.~\ref{fig:IRMSE_comp} shows component maps obtained in this case for LADMM, which were closely matched to the ADMM results (not shown due to space constraints, although spatial maps and $T_1$-$T_2$ spectra for ADMM were shown and discussed in detail in Ref.~\cite{Daeun2020}).  Fig.~\ref{fig:IRMSE_conv} compares the convergence characteristics of ADMM and LADMM for this case, where we still observe that the DFCS for LADMM converges much faster than for ADMM.  For example, the DFCS value obtained by ADMM after 15.0 hours of computation was achieved by LADMM in only 2.19 hours  (roughly a 6.8$\times$ improvement).  Note that this problem size is larger  than it was in the previous DR-CSI case, and the computation times are also longer. LADMM also used substantially less memory than ADMM in this case.  Peak memory usage was 43.5 MB for LADMM versus 95.6 MB for ADMM (roughly a 2.2$\times$ improvement).

\subsection{Multi-Echo Spin Echo $T_2$ Relaxometry}

The previous subsections demonstrated that LADMM could substantially accelerate computations in high-dimensional problems, where a 2D spectrum was estimated at every spatial location.  In this subsection, we evaluate a lower-dimensional problem where we desire to estimate a 1D $T_2$-relaxation spectrum from a series of multi-echo spin-echo data, while still using spatial regularization to improve the estimation results.  Specifically, we considered the in vivo human brain multicomponent $T_2$ dataset used in Fig.~11 of Ref.~\cite{Daeun2020}, where data acquisition involved $P=32$ $T_2$ encodings, the dictionary had $Q=300$ elements, and there were $N=4475$ voxels. The resulting 3D spectroscopic image was comprised of a 1D $T_2$ spectrum of size 300 at each spatial image location.

\begin{figure}[t]
\begin{center}
\includegraphics[scale=1]{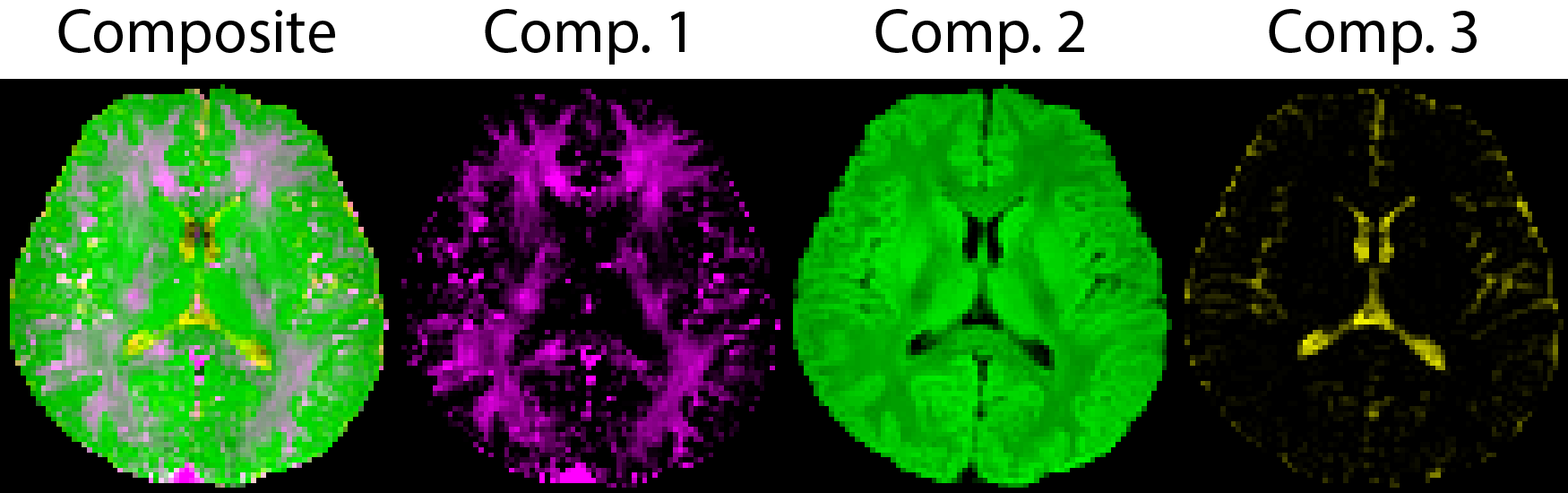}
\end{center}
	\caption{Component maps obtained for $T_2$ relaxometry data using LADMM.  The component maps were obtained by spectrally-integrating different regions of the 1D $T_2$ spectrum as described in Ref.~\cite{Daeun2020}.}
	\label{fig:T2_comp}
\end{figure}

Fig.~\ref{fig:T2_comp} shows component maps obtained in this case for LADMM, which were closely matched to the ADMM results (not shown due to space constraints, although spatial maps and $T_2$ spectra for ADMM were shown and discussed in detail in Ref.~\cite{Daeun2020}).  Notably, this type of 1D relaxometry experiment is less powerful than the 2D relaxometry experiment described in the previous subsection, as can be seen from the fact only 3 components are successfully resolved in this case, as compared to 6 components for RR-CSI. However, the 1D experiment has the benefit of requiring a much shorter acquisition than a 2D experiment, while also requiring less computational effort.

\begin{figure}[t]
	\centering
	\subfloat{\includegraphics[width=1.6in]{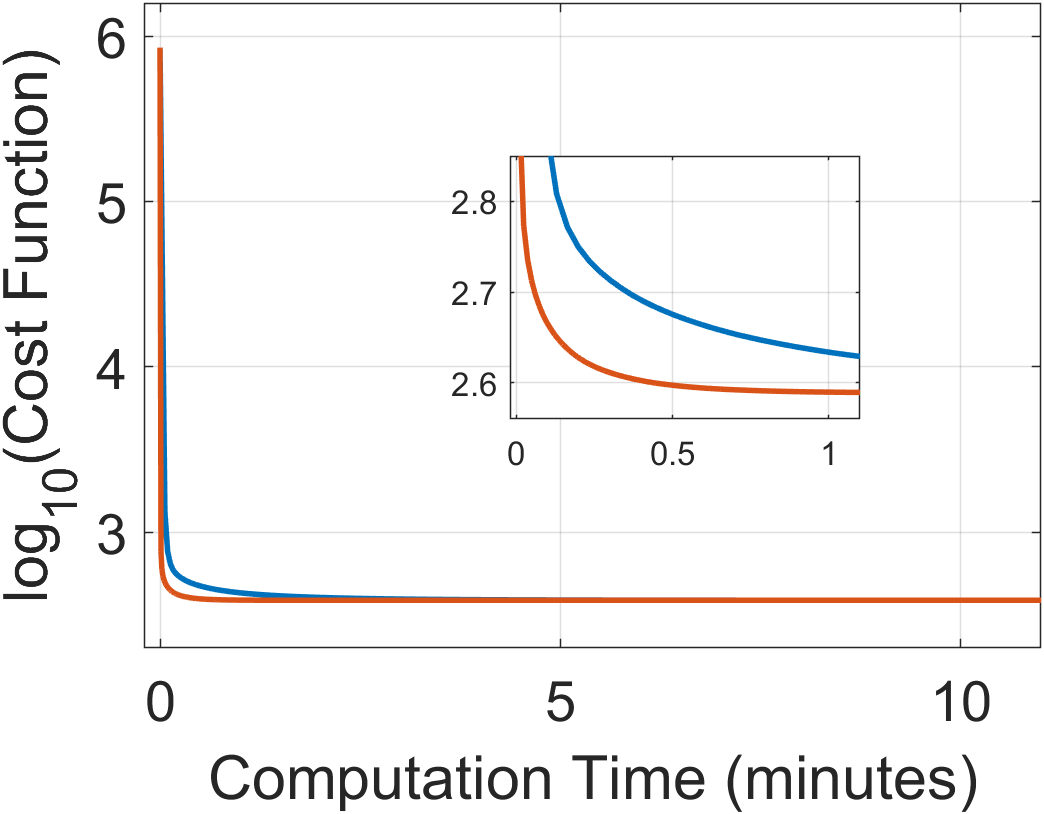}}
	\hfil
	\subfloat{\includegraphics[width=1.6in]{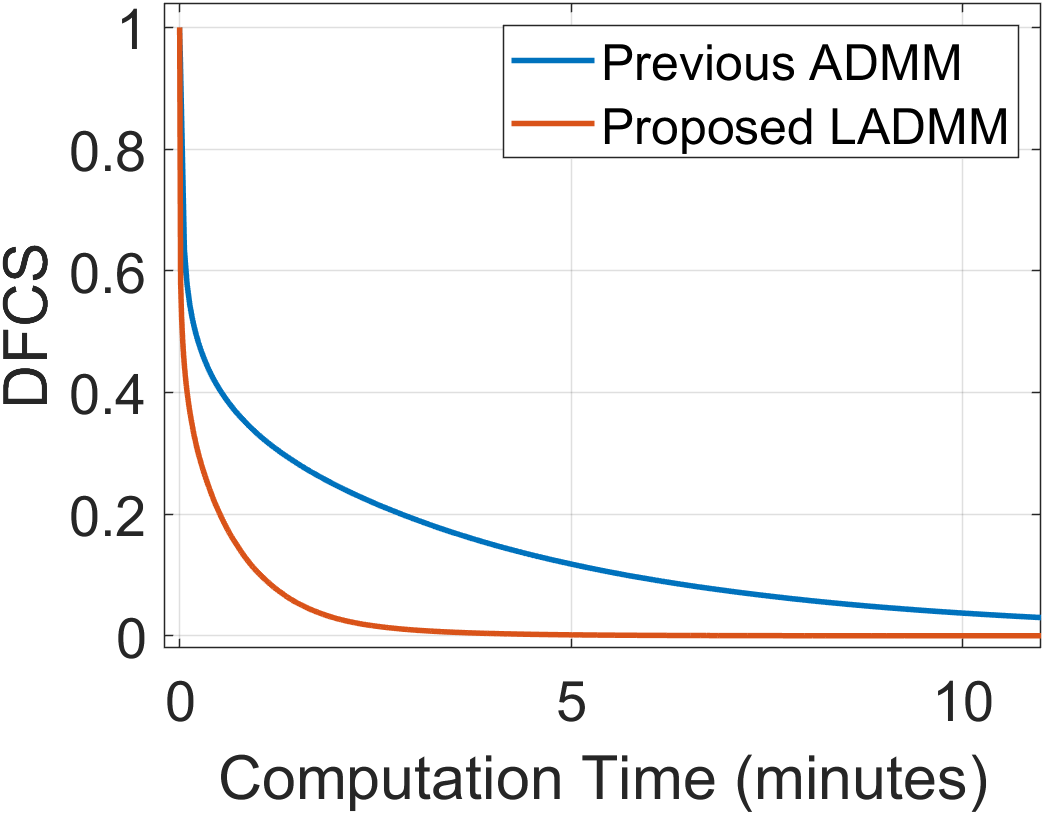}}
	\caption{Convergence of ADMM and LADMM for $T_2$ relaxometry data.}
	\label{fig:MSET2_conv}
\end{figure}

Fig.~\ref{fig:MSET2_conv} compares the convergence characteristics of ADMM and LADMM for this case, where we again observe that the DFCS for LADMM converges much faster than ADMM.  The DFCS value obtained by ADMM after 10.0 minutes of computation was achieved by LADMM in only 1.80 minutes (roughly a 5.6$\times$ improvement). Peak memory was also substantially smaller for LADMM (1.3 MB) compared to ADMM (19.1 MB), roughly a 14.9$\times$ improvement.

\begin{figure*}[t]
	\begin{center}
\includegraphics[scale=1]{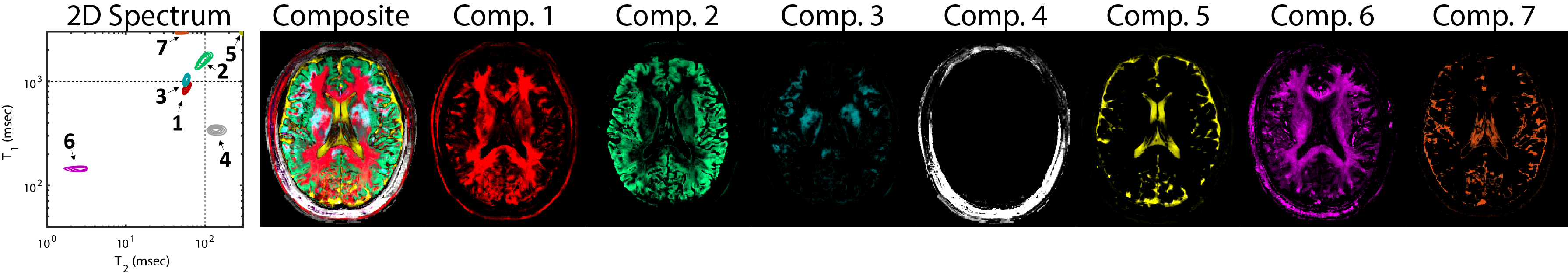}
	\end{center}
	\caption{Representative 2D spectra and component maps obtained for MRF data using LADMM. }
	\label{fig:MRF_comp}
\end{figure*}

\begin{figure}[t]
	\centering
	\subfloat{\includegraphics[width=1.6in]{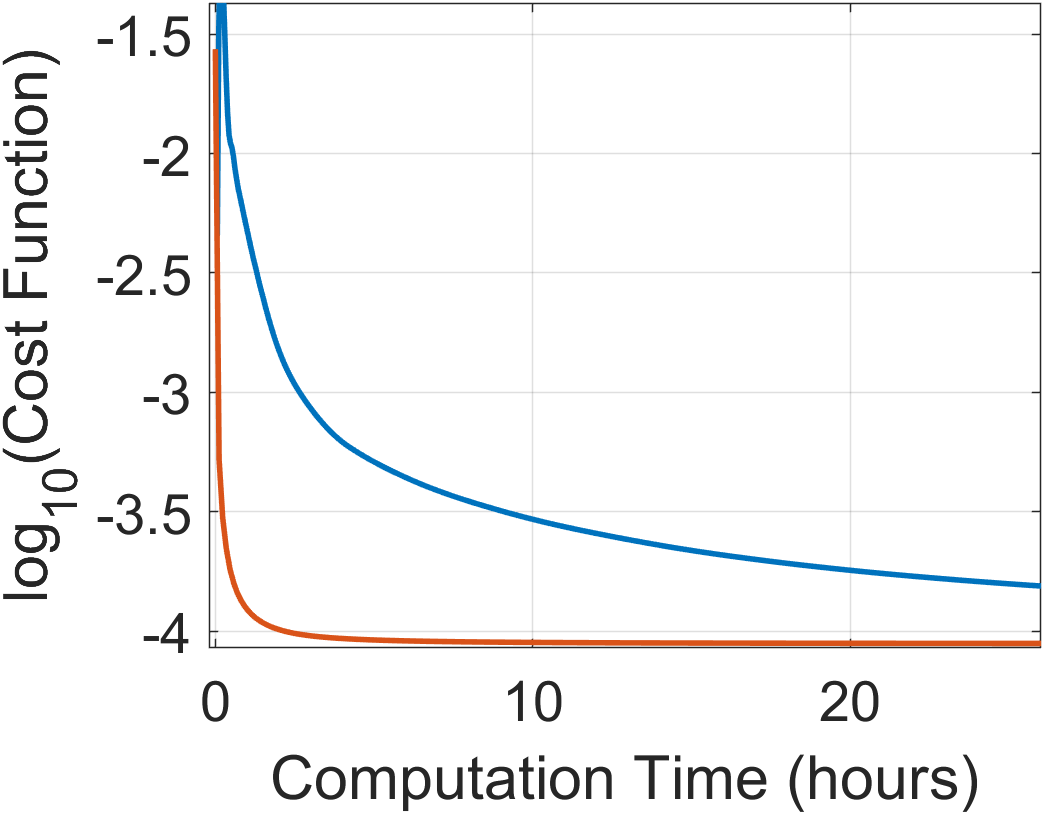}}
	\hfil
	\subfloat{\includegraphics[width=1.6in]{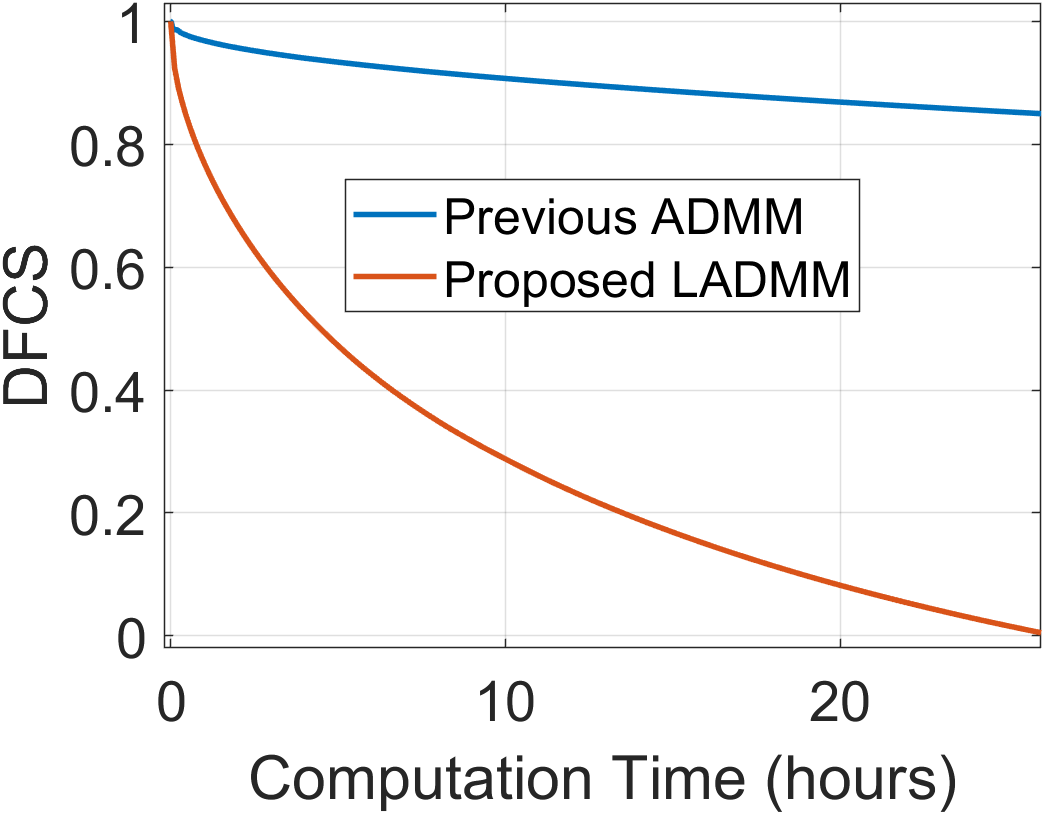}}
	\caption{Convergence of ADMM and LADMM for MRF data.}
	\label{fig:MRF_conv}
\end{figure}

\begin{figure}[t]
	\centering
\includegraphics[width=1.6in]{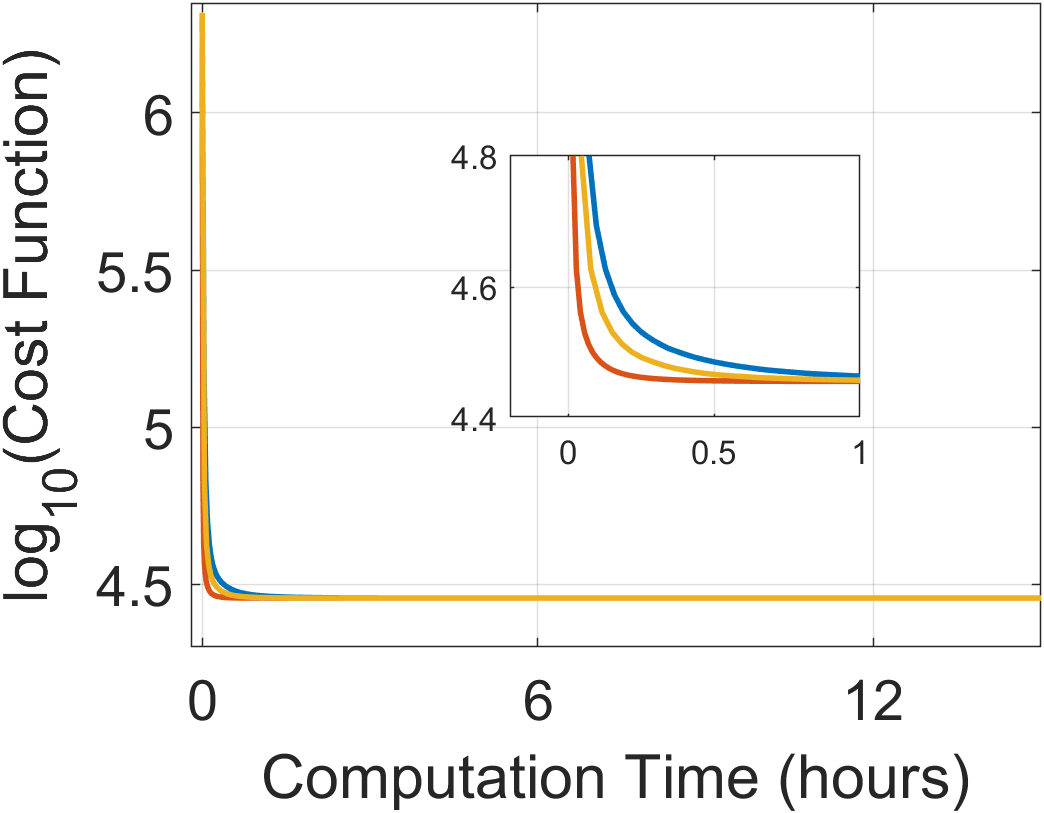}
	\hfil
	\includegraphics[width=1.6in]{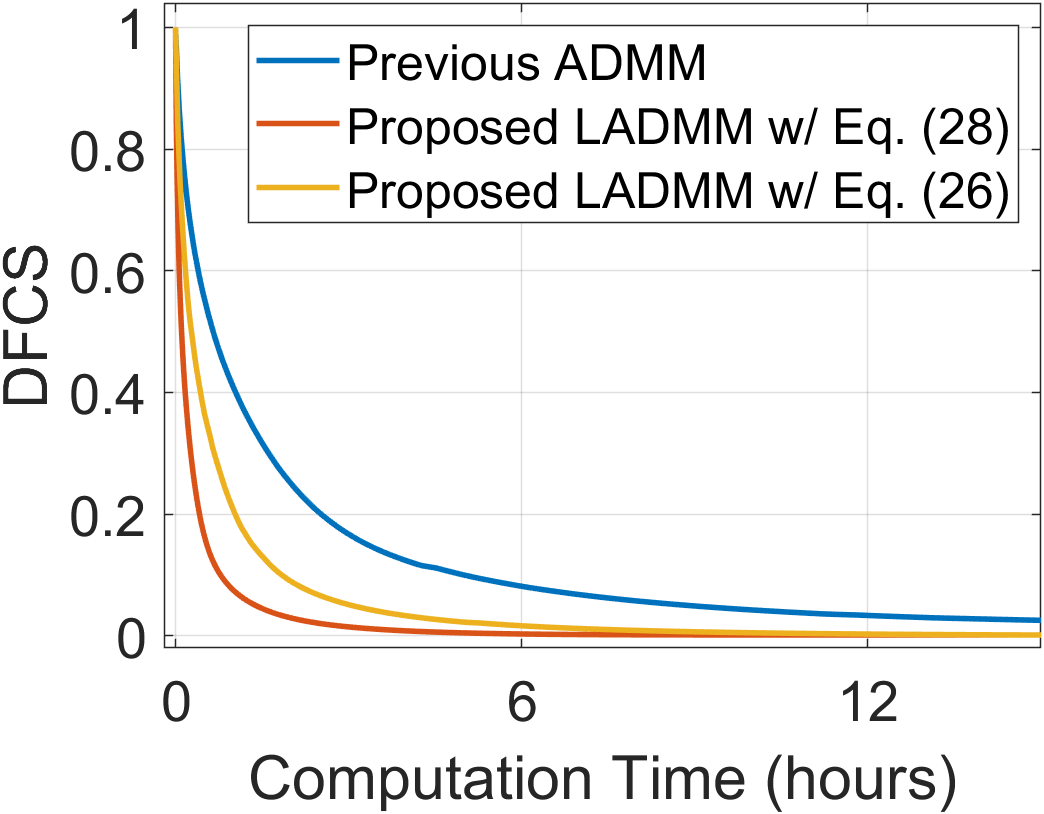}
	\caption{Convergence of LADMM using full-rank matrices (Eq.~\eqref{eq:DictMatxInv}) versus low-rank matrix approximations (Eq.~\eqref{eq:DictVecSVD}). }
	\label{fig:matx_inv_vs_svd}
\end{figure}

\subsection{MR Fingerprinting RR-CSI}
As a final test case, we considered estimation of $T_1$-$T_2$ spectroscopic images from data acquired using an MRF acquisition with a ViSTa preparation block to emphasize short $T_1$ components such as myelin water \cite{liu2022,Congyu2023}. A 2D slice of the in vivo human brain was acquired without acceleration ($32$ spiral interleaves to cover k-space at the Nyquist rate) with $1$ mm in-plane spatial resolution and $10$ mm slice thickness using a $3$T Siemens Prisma scanner. A total of $P = 540$ contrast-encoded images were acquired with varying sequence parameters.  To account for $B_1$ (flip angle) inhomogeneity, we acquired a $B_1$ map which was quantized to 61 different levels (uniformly spaced  from 0.7-1.3, where a value of 1.0 indicates a nominal flip angle).  Dictionaries were constructed using extended phase graph simulations for $(T_1,T_2)$ parameter  combinations corresponding to every possible combination of  101 different $T_1$ values (ranging from 100-3000 msec, spaced logarithmically) with 101 different $T_2$ values (ranging from 1-300 msec, spaced logarithmically) for each of the $B_1$ values ($Q=10201$), and different dictionaries were used for different spatial locations based on the $B_1$ map.   The resulting 4D spectroscopic image was comprised of a 2D $T_2$-$T_2$ spectrum of size 101$\times$101 at each of $N=24345$ spatial locations. 

Fig.~\ref{fig:MRF_comp} shows component maps obtained using LADMM for this case, where we were successfully able to resolve 7 anatomically plausible tissue components.   ADMM  results are not shown due to space constraints.  Unlike the other cases,  we observed some substantial differences between LADMM and ADMM in this case because ADMM did not converge in a reasonable amount of time, although we expect the results of ADMM and LADMM to be closely matched given sufficient time for ADMM to converge.  Fig.~\ref{fig:MRF_conv} compares convergence characteristics, where we again observe that the DFCS for LADMM converged substantially faster than for ADMM.  For example, the DFCS value obtained for ADMM after 25.0 hours was obtained by LADMM after only 0.460 hours (roughly a 54.4$\times$ improvement). Peak memory usage for LADMM (16.3 GB) was roughly 4.6$\times$ smaller than for ADMM (75.5 GB).

\section{Discussion}
\label{sec:disc}
The results in the previous section demonstrated that the new LADMM algorithm offers substantial advantages compared to the previous ADMM algorithm across a range of different problems.  While these improvements were primarily due to the problem simplifications offered by LADMM, the ability to use low-rank approximation (Eq.~\eqref{eq:DictVecSVD}) also contributed to computational efficiency.  The benefits of low-rank approximation for the RR-CSI data are illustrated in Fig.~\ref{fig:matx_inv_vs_svd}, where we observe that the use of low-rank approximation yields an obvious improvement in computation speed.

An interesting observation is that LADMM simplifies to ADMM if we take $\mathbf{P}=\mathbf{0}$ and $\mathbf{Q}=\mathbf{0}$. Since we have already taken $\mathbf{Q}=\mathbf{0}$ in our LADMM implementation, an ADMM variant of our new approach could be obtained by further setting $\mathbf{P}=\mathbf{0}$  (note that this would be a different ADMM implementation than the ones used in previous literature, since we rely on a different variable-splitting strategy as described in Sec.~\ref{sec:propm}).  However, it should be noted that the efficiency of solving the $\mathbf{z}$ subproblem (Eq.~\eqref{eq:z3}) in our LADMM implementation is dependent on our use of a nonzero $\mathbf{P}$ matrix. As discussed in Sec.~\ref{sec:z}, solving Eq.~\eqref{eq:z3} is nontrivial unless $\mathbf{P}$ is chosen carefully, which was one of the reasons that a different variable-splitting strategy was used in previous ADMM work.  In other words,  the additional flexibility of LADMM was a key ingredient to our success, since it opened the door to a powerful variable-splitting strategy that was not attractive with conventional ADMM.  

A limitation of our study is that we cannot definitively generalize our results to unseen datasets. While LADMM was empirically better than ADMM in all the cases we tried, and although we strongly suspect that this behavior will still hold true for similar unseen data, we have no theory that would allow us to definitively claim that LADMM will be better than ADMM for an unseen problem instance. This is a common limitation of all research based on empirical data. 

Notably, although LADMM was substantially faster than ADMM in all cases we tried, the computation speeds we achieved with LADMM (which ranged from couple of minutes to a couple of hours) may or may not be practically useful depending on the application.   In scientific studies, where data analysis is frequently performed weeks, months, or potentially even years after the original data collection, fast computations are convenient but may not be strictly necessary.  On the other hand, certain types of clinical applications may have much more urgency, and a couple hours of computation time may not be sufficient.  There are many avenues for further improvement in computation time, including better algorithms and/or the use of more advanced computation hardware.  

A feature of our LADMM approach (although not unique to LADMM, and also present for ADMM) is that the $\mathbf{f}$-subproblem is easily parallelized across different spatial locations, while the $\mathbf{z}$-subproblem is easily parallelized across different spectral positions.  This suggests that the proposed LADMM approach could be further accelerated in a parallel computing environment -- we did not explore that approach here, although this could be an interesting direction for future research.

In this work, we applied LADMM to solve an ill-posed inverse problem with nonnegativity constraints, quadratic data-fidelity constraints, and a quadratic spatial-roughness penalty. However, the principles we used in this scenario are quite general, and we anticipate that similar approaches may prove useful across a wide range of different optimization problems with similar structure.  While different problem formulations will likely require different approaches for variable splitting and linearization to achieve algorithmic efficiency, the previous LADMM literature \cite{Zhang2011,He2020,Fazel2013,Tao2020,Deng2016} should be a good starting point for researchers interested in applying LADMM to new objective functions.  

\section{Conclusion}
\label{sec:conc}
We proposed and evaluated an efficient algorithm, based on LADMM, for spatially-regularized partial volume component mapping from high-dimensional multiparametric MRI data.  The proposed approach was demonstrated to enable substantial computational improvements (ranging between roughly 3.1$\times$-54.4$\times$ acceleration for the cases we tried) across a range of different scenarios.  We expect that this reduction in computational complexity will make it  easier for the community to access the substantial benefits offered by spatial regularization in these kinds of problem settings.  Furthermore, the performance of LADMM in this scenario exceeded our expectations, and suggests the potential value of LADMM more broadly.

\bibliographystyle{IEEEtran}
\bibliography{./References}

\end{document}